\documentclass[final]{elsarticle}
\pdfoutput=1
\usepackage{hyperref}
\usepackage{lineno,color}
\journal{NIM-A}
\usepackage{amsmath}
\usepackage{graphicx}
\usepackage[left=2cm,right=2cm,top=2cm,bottom=2cm]{geometry}

\newcommand\aap{A\&A}
\newcommand\apjl{ApJ}

\def\procspie{\ref@jnl{Proc.~SPIE}} 

\usepackage{float}
\usepackage{eurosym}
\usepackage{subcaption}
\usepackage{multirow}

\biboptions{numbers}

\begin{document}

\begin{frontmatter}

\title{Side-On transition radiation detector: a detector prototype for TeV energy scale calibration of calorimeters in space}

\author[GXUaddress]{Bo Huang}
\author[GXUaddress]{Hongbang Liu \corref{cor}} 
\ead{liuhb@gxu.edu.cn}
\author[GXUaddress]{Xuefeng Huang}
\author[IHEPaddress]{Ming Xu}
\author[IHEPaddress]{Yongwei Dong}
\author[UCAS]{Xiaotong Wei}
\author[GXUaddress]{Xiwen Liu}
\author[GXUaddress]{Huanbo Feng}
\author[GXUaddress]{Qinhe Yang}
\author[GXUaddress]{Jianyu Gu}
\author[GXUaddress]{Shuai Chen}
\author[GXUaddress]{Xiaochuan Xie}
\author[GXUaddress]{Jin Zhang}
\author[GXUaddress]{Yongbo Huang}
\author[GXUaddress]{Enwei Liang \corref{cor}}
\ead{lew@gxu.edu.cn}

\cortext[cor]{Corresponding author}

\address[GXUaddress]{Guangxi Key Laboratory for Relativistic Astrophysics, School of Physical Science and Technology, Guangxi University, Nanning, China}
\address[IHEPaddress]{Key Laboratory of Particle Astrophysics, Institute of High Energy Physics, Chinese Academy of Sciences, Beijing, China}
\address[UCAS]{School of Physical Science, University of Chinese Academy of Sciences, Beijing, China}

\begin{abstract}

Transition Radiation (TR) plays an important role in particle identification in high-energy physics and its characteristics provide a feasible method of energy calibration in the energy range up to 10 TeV, which is of interest for dark matter searches in cosmic rays. In a Transition Radiation Detector (TRD), the TR signal is superimposed onto the ionization energy loss signal induced by incident charged particles. In order to make the TR signal stand out from the background of ionization energy loss in a significant way, we optimized both the radiators and the detector. We have designed a new prototype of regular radiator optimized for a maximal TR photon yield, combined with the Side-On TRD which is supposed to improve the detection efficiency of TR. We started a test beam experiment with the Side-On TRD at Conseil Europ\'{e}en pour la Recherche Nucl\'{e}aire (CERN), and found that the experimental data is consistent with the simulation results.

\end{abstract}


\end{frontmatter}

\section{Introduction}
\label{sec:intro}

Dark matter is one of the fundamental challenges of astronomy in this century, and its detection is also attempted via the observations of the spectra of electrons and high energy cosmic rays. The current experiments, i.e. FERMI \citep{aa2009}, AMS-02 \citep{ma2014}, PAMELA \citep{oa2011}, ATIC \citep{jc2008}, CALET \citep{efm2017,ktz2018}, DAMPE \citep{wek2017}, show that dark matter may exist, and anomalies in the measured electron and positron spectra may be caused by dark matter. However, the experimental accuracy is still limited and the possible contribution from other astrophysical processes is still unresolved. Larger statistics need to be accumulated to analyze the multi TeV energy region in the spectra. In order to measure with precision high-energy cosmic-ray particles absorbed by a calorimeter, the calibration of the calorimeter is very important.

To evaluate the relationship between the energy deposited by incident particles and the resulting signal of the calorimeter, energy calibration is required. Precise TRD's designed to measure the Lorentz factor of cosmic rays were successfully used in the CRN \citep{msm1991} and TRACER \citep{ob2011} experiments, and the cross-calibration of a calorimeter and a TRD in-orbit was performed by the balloon-borne experiment CREAM \citep{ahn2008}. Moreover, the CREAM experiment exploited the cross-calibration in flight between calorimeter and TRD \citep{mae2010}. Extensive and innovative R\&D of TRDs for energy measurements of cosmic-ray nuclei at high Lorentz factors were also carried out in the 2000's \citep{wak2004,cas2004,che2004}. Currently, the energy scale can be calibrated by accelerated particle beams below 400 GeV at CERN. In the higher energy region up to TeV an energy calibration of calorimeters can be performed in-flight by using TRD with cosmic rays. 

Transition radiation, first predicted by Ginzberg and Frank \citep{vi1946}, occurs when a charged particle passes through an interface between media with different dielectric constants, or more generally through inhomogeneous media. The theory was refined by Cherry \citep{mlc1974}, Fabjan \citep{cw1975} and Artru \citep{xa1975}. TR is generated at about $\gamma\sim10^{3}$ ($\gamma$ is the Lorentz factor of the incident charged particle), reach a saturation at about $\gamma\sim10^{4}$, and its intensity is proportional to $\gamma$ \citep{xa1975}. Actually, TRDs detect the intensity of TR, from which $\gamma$ of the incident charged particle can be inferred. Combining the Lorentz factor with the particle energy obtained by the calorimeter, the particle mass can be identified. On the other hand, we can exploit the TR measurement to calculate the particle energy, which can be compared with the energy registered in the calorimeter to derive the energy calibration. The calorimeter of HERD (High Energy cosmic Radiation Detection facility) has been tested at the CERN SPS in November 2015 \citep{dong2016}. Since the energy interval where protons emit TR is approximately 1-10 TeV, this method provides a feasible energy calibration method for this calorimeter at the TeV energy scale. 

The main difficulty in the practical application of TR radiation is its small yield and the small angle of emission of the TR photons with respect to the direction of the charged particle. The small radiation yield results in a weak signal; the small emission angle of TR photons causes the signal generated by TR to overlap with the ionization energy loss \citep{dh1996, ca2008} signal induced by the incident charged particles. In order to detect a significant TR signal on top of the background of ionization energy loss, it is essential to optimize both the radiator and the TRD.

Radiators are mainly divided into two categories: regular and irregular. The regular radiators are commonly constructed using spacers of equal thickness and regular foils, the irregular radiators consist of foams or fiber materials. The regular radiators can produce a larger number of photons per charged particle and boundary by exploiting coherent interference, whereas the mean number of photons is lower in case of irregular radiators due to the missing coherent interference and/or the smaller structure size in foams or fiber materials \citep{bc2014}. In this work, we developed a simulation based on GEANT4 \citep{ja2016} to optimize the regular radiators with the aim to increase the yield of TR photons, and find a new, cost efective method of their production. It is described in Section \ref{sec:radiator} .

TRD is a device for charged particle identification. Some experiments like ALICE \citep{ac2018}, CBM \citep{mp2013}, AMS \citep{ts2006, pd2006} and the forthcoming HERD \citep{dong2019overall}, employ TRDs consisting of two different parts: a radiator, in which a charged particle creates TR, and a detection part consisting of a gas chamber filled with a gas. In order to obtain a high detection efficiency for the signal, the working gas thickness should be adjusted appropriately. In Section \ref{sec:TRD} of this paper we describe a new prototype of TRD called Side-On TRD.

After the first observation of TR in the optical region by Goldsmith and Jelley \citep{pj1959}, TRD was used for indentification of charged paticles. TR also provide a feasible method of energy calibration due to its properties. We compare the simulation results with the experimental data from test beam experiment at CERN SPS and see that they are in good agreement and it is discussed in Section \ref{sec:EC} .

Finally, we give discussions and summarize the main points of this work in Section \ref{sec:CON} .

\section{Simulation of regular radiators}
\label{sec:radiator}

The theoretical expression for the spectral intensity of transition radiation is derived in Ref. \citep{le1984}. However, in order to start this work, a detailed review of the fundamental derivation is not necessary. Here we summarize the theory step by step from single interface, single foil (double interfaces) to multi-layers radiators. The complete physics process for TR production is explained in the \ref{app:TR} . Moreover, the calculation of TR produced by irregular radiators is discussed in Ref. \citep{gmg1974,gmg1975,cwf1977}.

The TR emittance from a single surface can be described as
\begin{eqnarray}\label{eq1}
\biggl(\frac{d^{2}W}{d\omega d\theta} \biggr) _{interface}=\frac{2\alpha\theta^{3}}{\pi}\biggl(\frac{1}{\gamma^{-2}+\theta^{2}+\xi_{1}^{2}}-\frac{1}{\gamma^{-2}+\theta^{2}+\xi_{2}^{2}}\biggr)^{2}
\end{eqnarray}
where $\gamma$ is the particle Lorentz-factor, $\alpha=e^{2}/\hbar c\approx 1/137$ is the fine structure constant, $\theta$ is the emission angle of TR photons with respect to the direction of the charged particle, $\xi^2_{i}=\omega_{P,i}^{2}/\omega^{2}$, and $\omega_{P,i}$ is the plasma frequency of the foil or gap materials, respectively.

The TR emittance from a single foil (two interfaces) can be expressed as
\begin{eqnarray}\label{eq2}
\biggl(\frac{d^{2}W}{d\omega d\theta}\biggr)_{singlefoil}=\biggl(\frac{d^{2}W}{d\omega d\theta}\biggr)_{interface}\times 4\sin^{2}{\frac{\varphi_{1}}{2}}
\end{eqnarray}
$4\sin^{2}{\varphi_{1}/2}$ is the interference factor, $\varphi_{i}\approx {l_{i}}/{D_{i}}$. As for multi-layers, the interference effect should be taken into account. Thus, the TR spectrum resulting from a stack of $N$ regular foils of thickness $l_{1}$ seperated by distance $l_{2}$ can be described as

\begin{eqnarray}\label{eq3}
\biggl(\frac{d^{2}W}{d\omega d\theta}\biggr)_{stack}=\biggl(\frac{d^{2}W}{d\omega d\theta}\biggr)_{singlefoil}\times\frac{\sin^{2}[N(l_{1}/D_{1}+l_{2}/D_{2})]}{\sin^{2}(l_{1}/D_{1}+l_{2}/D_{2})}
\end{eqnarray}
$D_{i}$ is defined as the formation zone. Different materials have different formation zones, as it is shown in detail in \ref{app:TR} .

On the other hand, the losses due to absorption effect of the foils and gaps must be taken into account, so the average value of Eq.\ref{eq3} is given by:
\begin{eqnarray}\label{eq4}
\biggl(\frac{d^{2}W}{d\omega d\theta}\biggr)_{foil}=\biggl(\frac{d^{2}W}{d\omega d\theta}\biggr)_{stack}\times\exp\biggl(\frac{\sigma(1-N)}{2}\biggr)\frac{\sin^{2}(N\phi_{12}/2)+\sinh^{2}(N\sigma/4)}{\sin^{2}(\phi_{12}/2)+\sinh^{2}(\sigma/4)}
\end{eqnarray}
where $\phi_{12}=\phi_{1}+\phi_{2}$ is the phase difference with $\phi_{i}=(\gamma^{-2}+\theta^{2}+\xi_{i}^{2})\omega l_{i}/2$. The absorption cross section of the radiator materials is $\sigma=\sigma_{1}+\sigma_{2}=\mu_{1}l_{1}+\mu_{2}l_{2}$ (foil + gap), $\mu_{1,2}$ being the absorption coefficient of the different media \citep{aj2001}. 

From Eq.\ref{eq4}, it can be noticed that the TR yield depends on the absorption coefficient of materials, thickness of foil and gap, number of layers and incident particle Lorentz-factor. 

Usually radiators are devices used for producing TR photons emitted at periodic boundaries between different materials. The radiator is composed of two kinds of dielectric materials alternately arranged, and its structure is a multi-layer stack containing a large number of interfaces. The incident particles pass through these interfaces, generating at each interface forward TR photons (X-rays), which are coherent and superimposed with one another, and finally get out from the stack structure. Since the TR photon signal is superimposed onto the ionization signal of the charged particle, one method to distinguish the two signals in a more significant way is to optimize the TR yield of the radiators. Thus, it is essential to understand the influence of radiator parameters like materials, foil thickness, gap thickness and number of foils. The following simulation of Side-On TRD is based on GEANT4 \citep{ja2016}, and it uses the tool called RADIATOR \citep{pv2004} to optimize regular radiators.

\subsection{Radiator parameters}

\subsubsection{Materials}

The radiator has a self-absorption effect \citep{Goulon1982On} on TR photons, which affects the detection of TR photons. Different materials have different plasma frequencies and self-absorption coefficients under the same conditions of pressure and temperature. The properties and attenuation coefficients of several common foil and gap materials are shown in the \ref{app:xcom}.

In the actual case, since multi-foils are necessary, the material must be chosen with an X-ray mass attenuation coefficient as small as possible, in order to reduce the absorption of TR photons in the process of forward propagation. Li, Be, CH$_{2}$ (PP or PE) and mylar have been used as the radiator foil material in the past. According to \ref{app:xcom} , Li and Be are good foil materials, but they need to be kept in the inert gas due to their chemical activity. Therefore, PP is selected as the foil material here. H and He would be the preferred choice for the gap material, but for reasons like airtightness, complexity and cost of production, most commonly air is used. Thus, the following study is based on PP ($l_{1}=20\mu m$) as foil material and air ($l_{2}=800\mu m$) as gap material with $N=300$ foils.

\subsubsection{Foil thickness}

\begin{figure*}[ht]
	\centering
	\includegraphics[clip=true,trim=0cm 0cm 0cm 0cm,width= 0.45\textwidth]{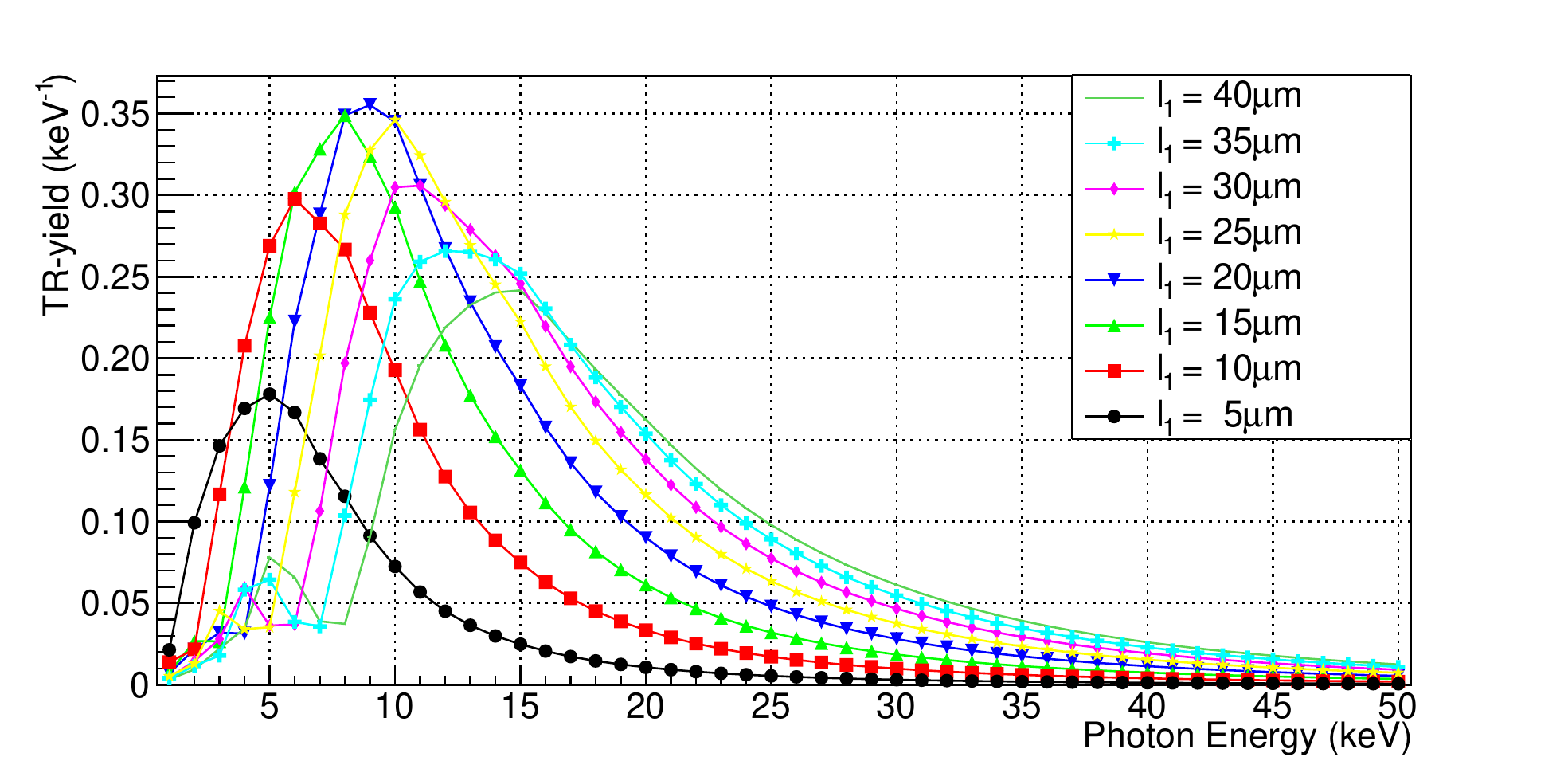}
	\includegraphics[clip=true,trim=0cm 0cm 0cm 0cm,width= 0.45\textwidth]{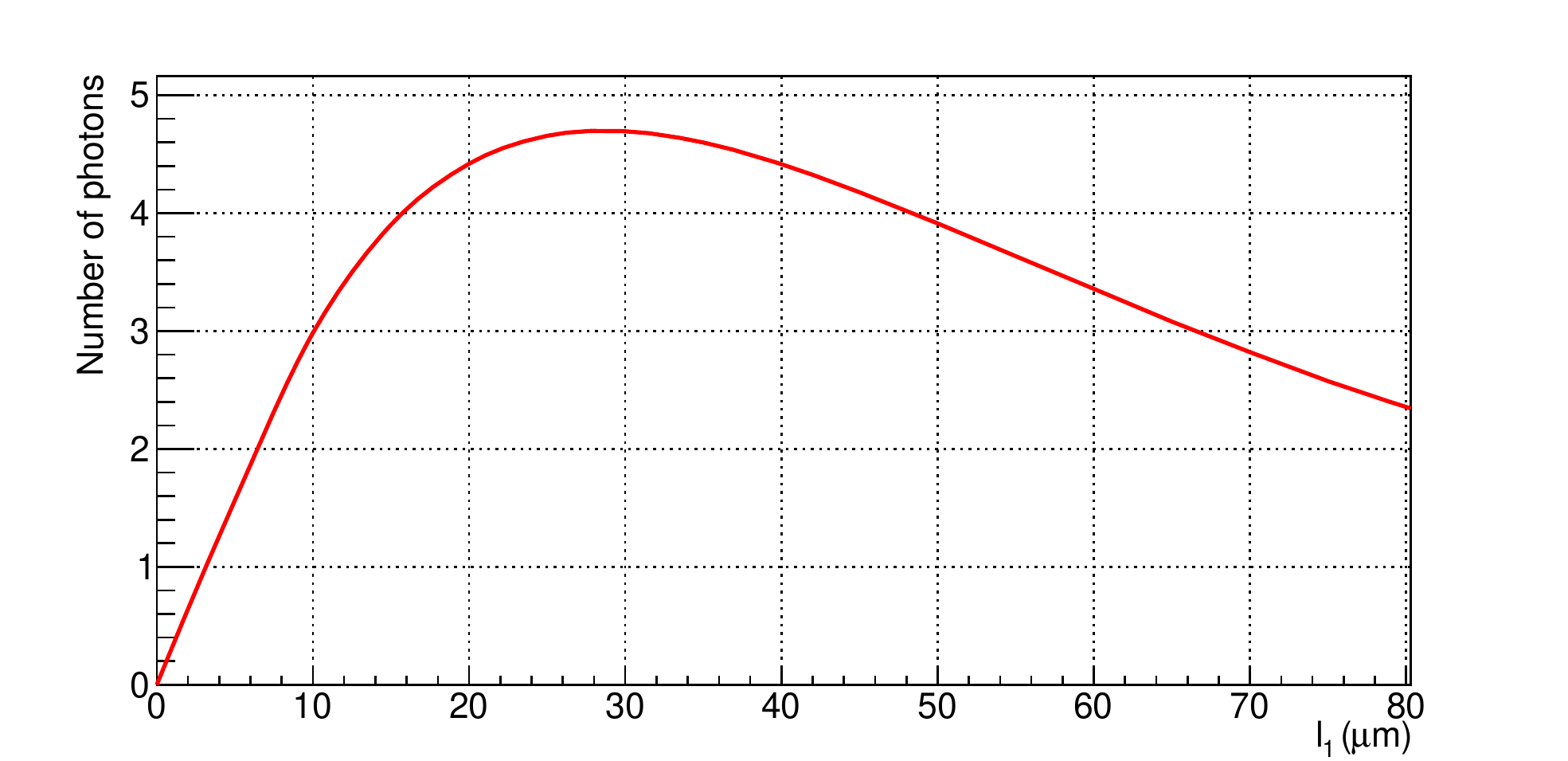}
	\caption{TR photon energy spectra for different foil thicknesses $l_{1}$ (left), and number of photons as a function of $l_{1}$ (right), for a radiator consisting of 300 PP foils with a fixed gap thickness of $l_{2}=800\mu m$ crossed by 2.5 GeV/c electrons. In the left-hand plot, solid line: $l_{1}=40 \mu m$; cross: $l_{1}=35 \mu m$; rhombus: $l_{1}=30 \mu m$; pentagram: $l_{1}=25 \mu m$; reversed triangle: $l_{1}=20 \mu m$; triangle: $l_{1}=15 \mu m$; square: $l_{1}=10 \mu m$; circle: $l_{1}=5 \mu m$.}
	\label{fig:foil}
\end{figure*}

The TR-spectrum and the total number of photons as a function of the foil thickness $l_{1}$ are shown in Fig.\ref{fig:foil}, for a radiator based on PP foils ($N=300$) and air gap ($l_{2}=800\mu m$) with 2.5 GeV/c incident electrons. The position of the TR-spectrum maximum, according to Eq.\ref{eqA14}, is shifted to higher energies as $l_{1}$ increases and the maximum is reached for a foil thickness between 15 and 25$\mu m$ at a photon energy of 8-10 keV (left figure). The total number of TR-photons is maximal for a foil thickness of 20-35$\mu m$ (right figure). Both, the height of the TR spectrum and the TR yield, decrease after the peak due to the increase of foil self-absorption.

\subsubsection{Gap thickness}

\begin{figure*}[ht]
	\centering
	\includegraphics[clip=true,trim=0cm 0cm 0cm 0cm,width= 0.45\textwidth]{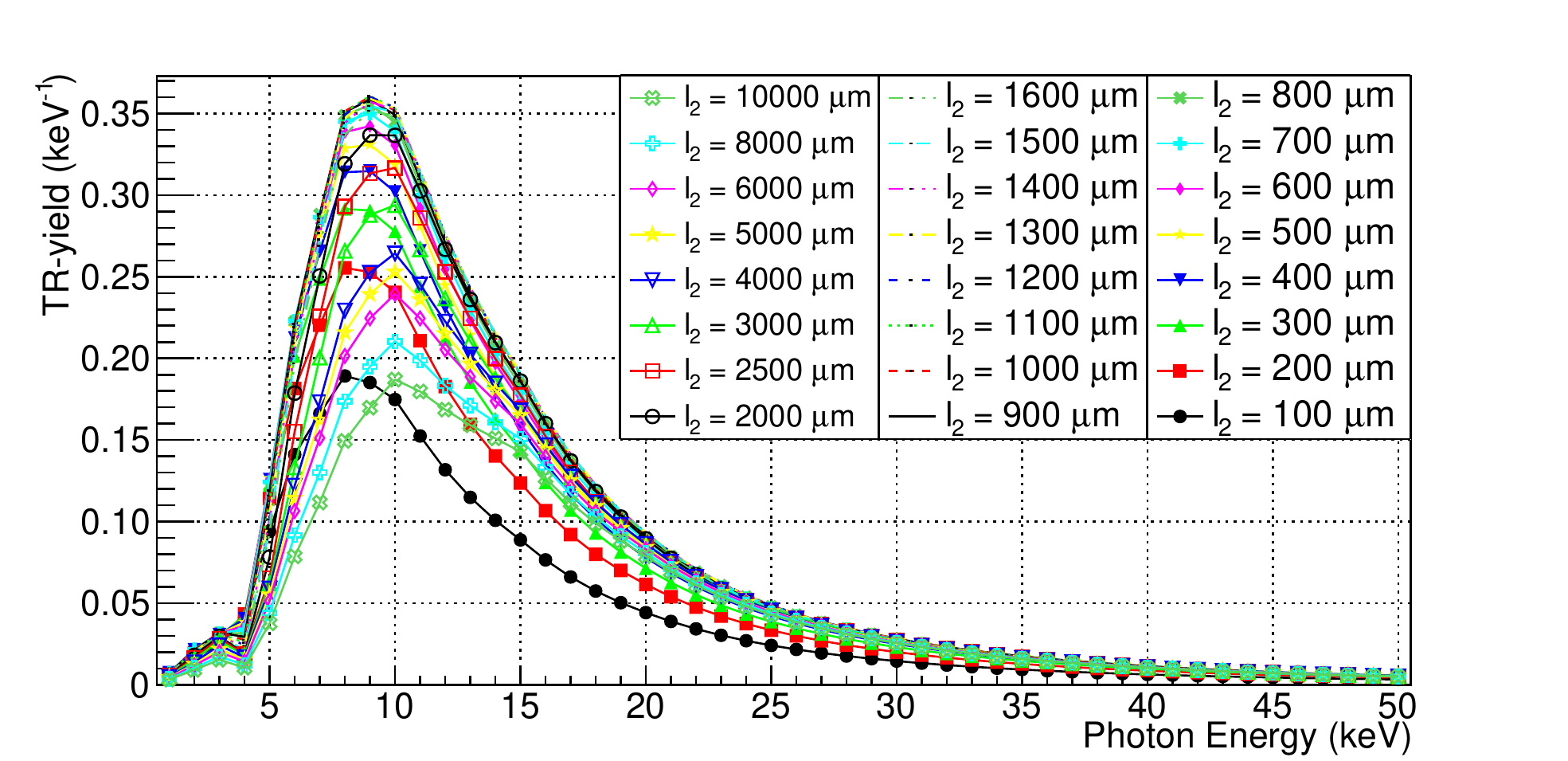}
	\includegraphics[clip=true,trim=0cm 0cm 0cm 0cm,width= 0.45\textwidth]{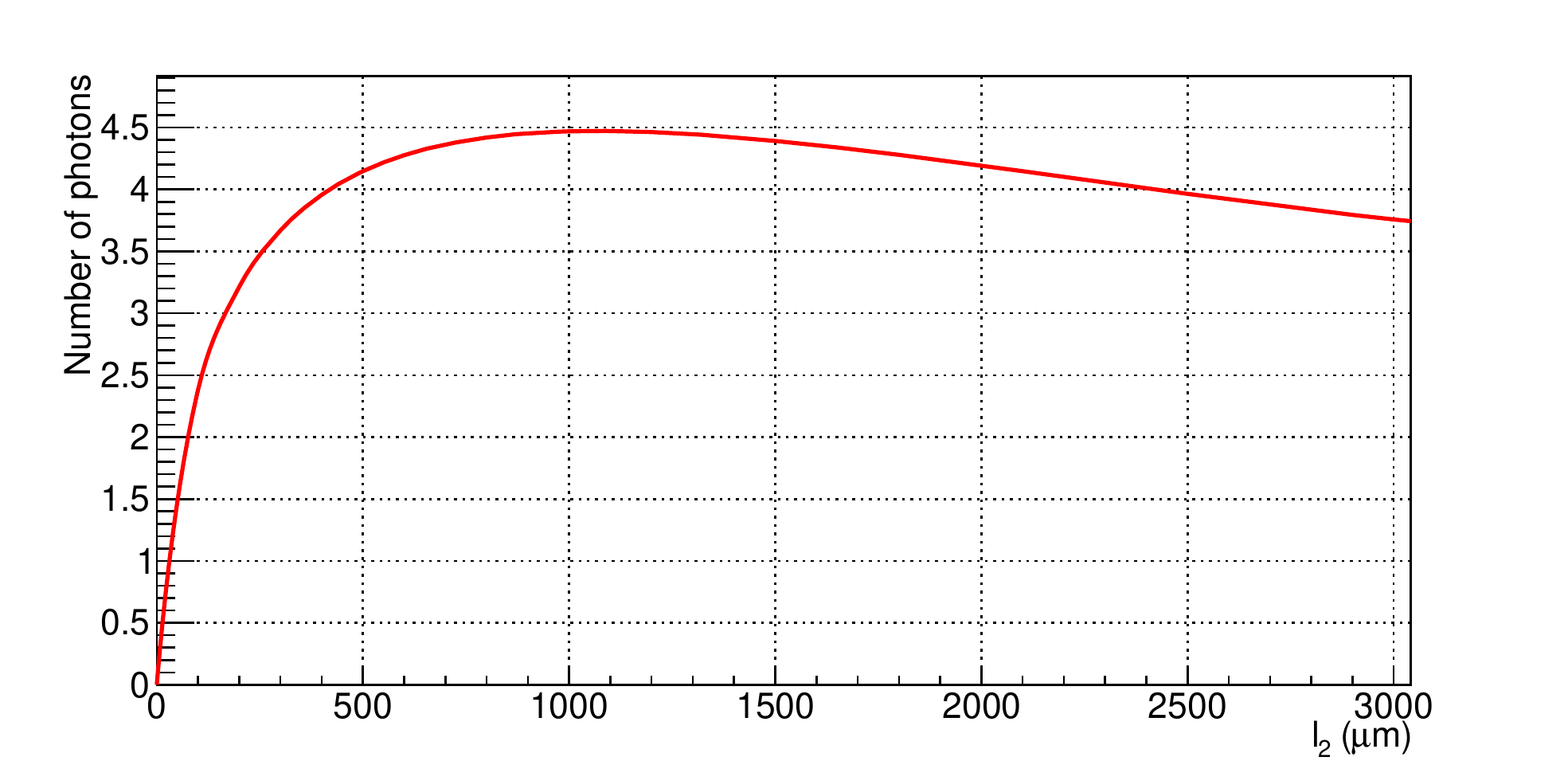}
	\caption{(colour online) TR photon energy spectra for different gap thicknesses $l_{2}$ (left), and number of photons as a function of $l_{2}$ (right), for a radiator consisting of 300 PP foils ($l_{1}=20\mu m$) crossed by 2.5 GeV/c electrons. In the left-hand plot, solid multiplication sign: $l_{2}=800 \mu m$, solid cross: $l_{2}=700 \mu m$, solid rhombus: $l_{2}=600 \mu m$, solid pentagram: $l_{2}=500 \mu m$, solid reversed triangle: $l_{2}=400 \mu m$, solid triangle: $l_{2}=300 \mu m$, solid square: $l_{2}=200 \mu m$, solid circle: $l_{2}=100 \mu m$; hollow multiplication sign: $l_{2}=10000 \mu m$; hollow cross: $l_{2}=8000 \mu m$, hollow rhombus: $l_{2}=6000 \mu m$, hollow pentagram: $l_{2}=5000 \mu m$, hollow reversed triangle: $l_{2}=4000 \mu m$, hollow triangle: $l_{2}=3000 \mu m$, hollow square: $l_{2}=2500 \mu m$, hollow circle: $l_{2}=2000 \mu m$; different line styles indicate the gap thickness from 900 to 1600 $\mu m$, respectively, and they almost overlap each other at the height of the TR spectrum.}
	\label{fig:gap}
\end{figure*}

The influence of the gap thickness $l_{2}$ on the energy spectrum of TR photons and the number of TR photons is shown in Fig.\ref{fig:gap}. The maximum of the TR spectrum is shifted slightly to higher energies as $l_{2}$ increases (left figure). And the height of the TR spectrum (left figure) and the number of TR photons (right figure) are maximal for a gap thickness between 800 and 1600$\mu m$. As $l_{2}$ increases, both of them decrease slightly due to the weak self-absorption in the gap.

\subsubsection{Number of foils}

\begin{figure*}[ht]
	\centering
	\includegraphics[clip=true,trim=0cm 0cm 0cm 0cm,width=0.45\textwidth]{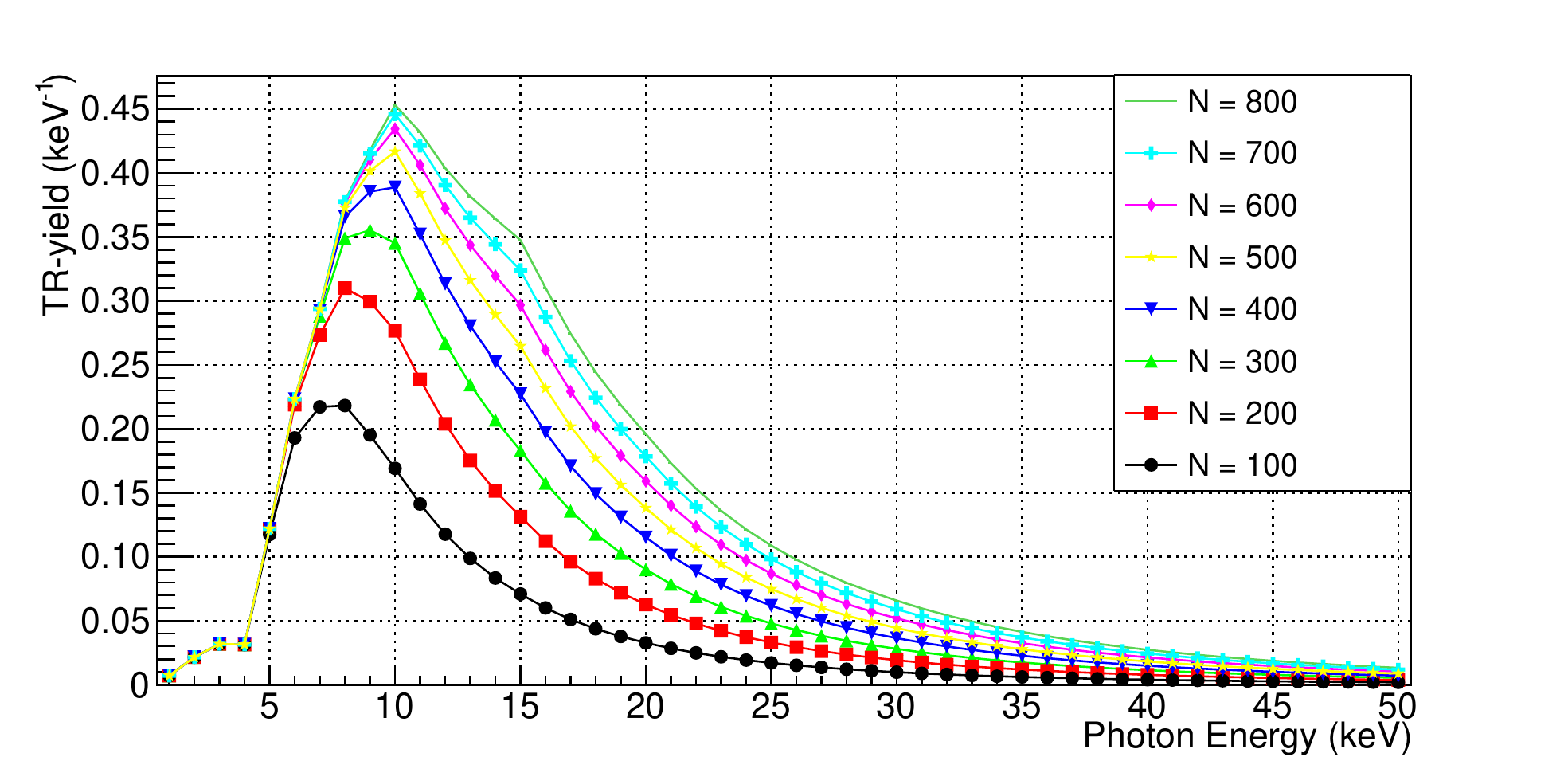}
	\includegraphics[clip=true,trim=0cm 0cm 0cm 0cm,width=0.45\textwidth]{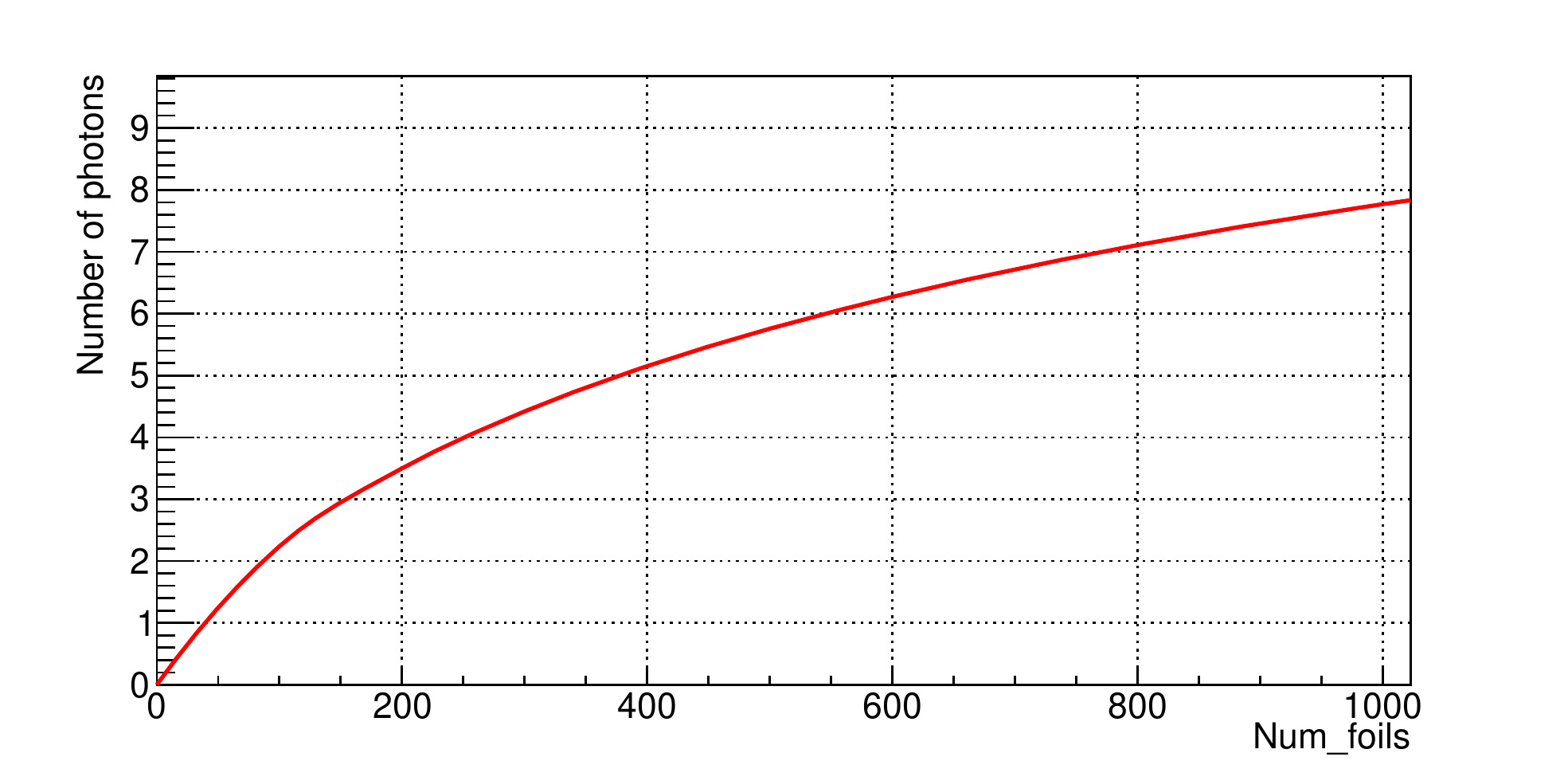}
	\caption{TR spectrum as a function of photon energy for different number of foils $N$ (left), and number of photons as a function of $N$ (right), for a radiator consisting of PP foil ($l_{1}=20\mu m$) and air gap ($l_{2}=800\mu m$) crossed by 2.5 GeV/c electrons. In the left-hand plot, solid line: $N$=800; cross: $N$=700; rhombus: $N$=600; pentagram: $N$=500; reversed triangle: $N$=400; triangle: $N$=300; square: $N$=200; circle: $N$=100.} 
	\label{fig:Nfoils}
\end{figure*}

The influence of the number of foils on the TR-spectrum and TR-yield is shown in Fig.\ref{fig:Nfoils}. As the number of foils increase, both, the height of the TR spectrum and the TR yield increase. However, the growth rate of TR yield decreases as the mass and volume of the radiator also rapidly rise with increasing values of $N$. That should be taken into account when designing the radiators.

\subsubsection{Lorentz factor}

\begin{figure*}
	\centering
	\includegraphics[clip=true,trim=0cm 0cm 0cm 0cm,width=0.45\textwidth]{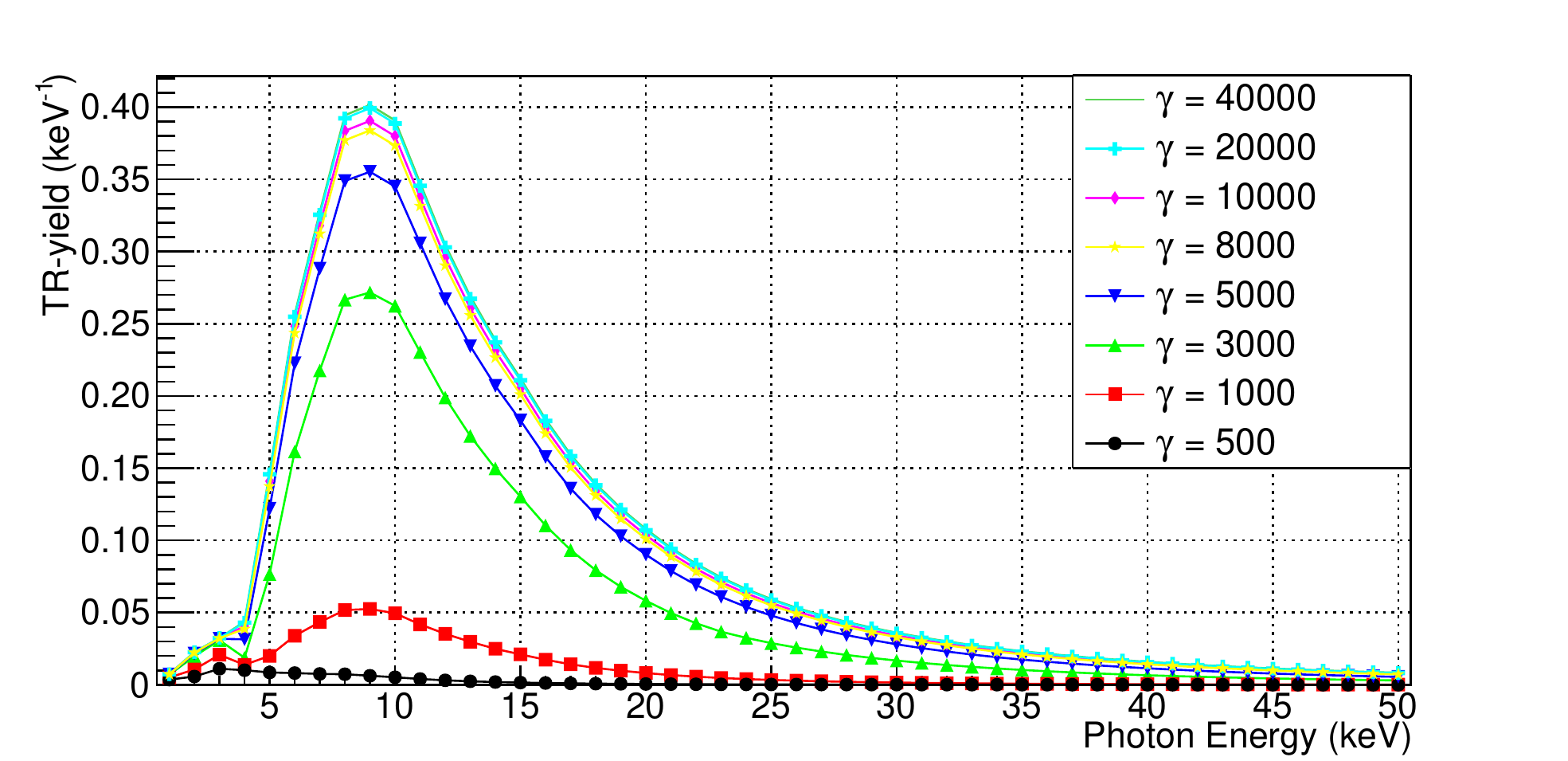}
	\includegraphics[clip=true,trim=0cm 0cm 0cm 0cm,width=0.45\textwidth]{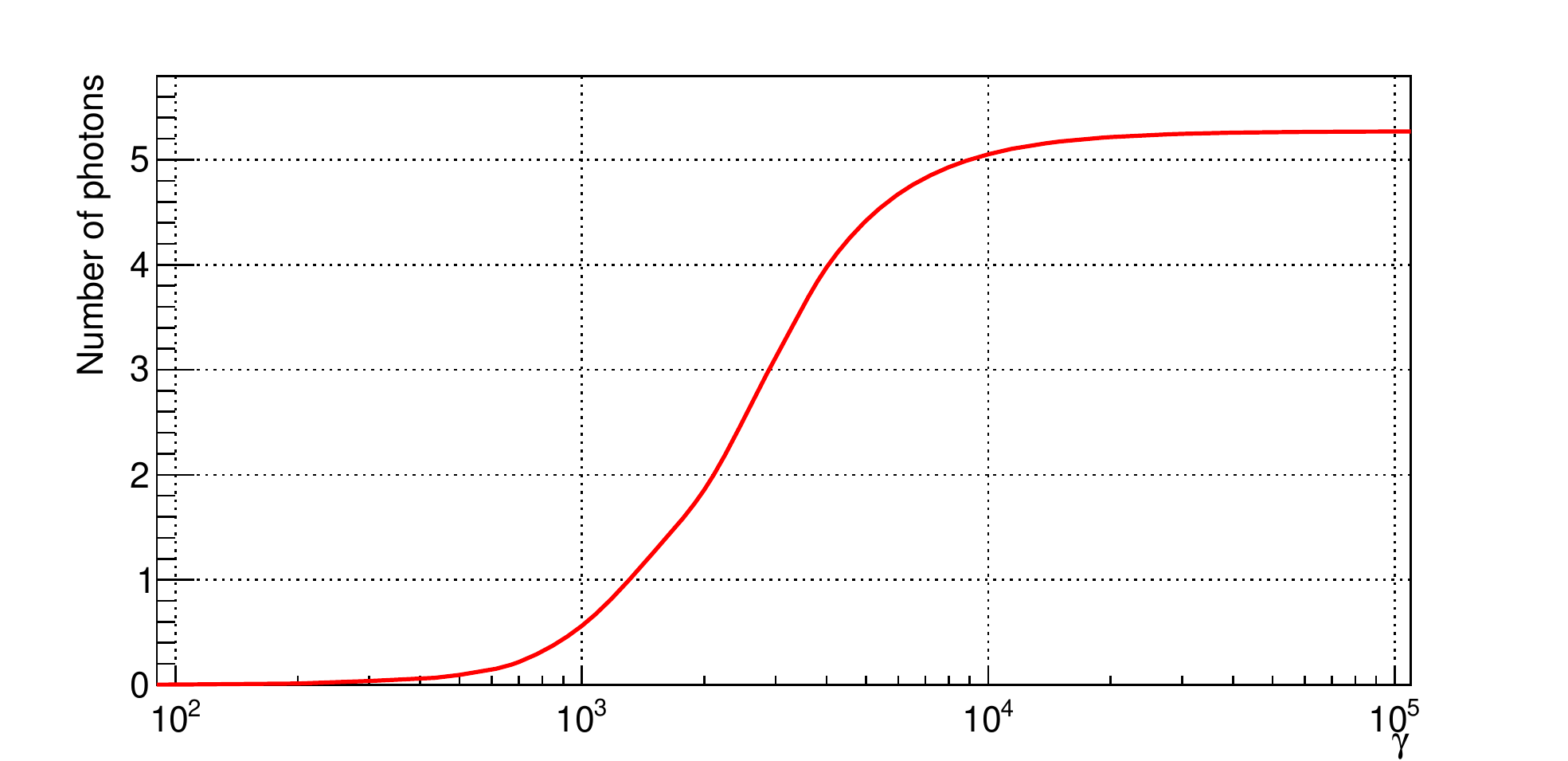}
	\caption{TR photon energy spectra for different Lorentz factors $\gamma$ (left), and the number of TR photons as a function of $\gamma$ (right), for a radiator consisting of $N=300$ PP foils ($l_{1}=20\mu m$)  and air gap ($l_{2}=800\mu m$). In the left-hand plot, solid line: $\gamma$=40000; cross: $\gamma$=20000; rhombus: $\gamma$=10000; pentagram: $\gamma$=8000; reversed triangle: $\gamma$=5000; triangle: $\gamma$=3000; square: $\gamma$=1000; circle: $\gamma$=500.} 
	\label{fig:Lfactor}
\end{figure*}

The relations between the Lorentz factor of the incident particle and th TR-spectrum and the TR yield are shown in Fig.\ref{fig:Lfactor}. The maximal height of the TR spectra is reached for a radiator with foil ($l_{1}=20\mu m, N=300$) and air gap ($l_{2}=800\mu m$) at a photon energy of 8-10 keV. As $\gamma$ increases, the maximum of the TR spectrum and, consequently, also the TR yield increases for $\gamma\textgreater10^{3}$ and reaches a saturation at $\gamma\sim10^{4}$. 

~\\
It can be concluded from the above optimization of radiator parameters, that foil material, foil thickness, gap material, gap thickness and number of foils all have an influence on the TR production. Different materials have different self-absorption of TR photons, so the materials with low self-absorption and stable properties should be selected as the foil and gap material of the radiator. The foil thickness has a great influence on the peak position of the TR spectrum, while the gap thickness has little influence on it. In other words, selecting the appropriate foil and gap thickness can shift the maximum of the TR-spectrum towards the high energy region. Increasing the number of layers can increase the maximum of TR-spectrum and the total number of TR photons, however, as the number of layers increases, the corresponding growth rate of the the number of photons gradually decreases, while the volume and mass of the radiator also increase rapidly.

\subsection{Regular radiators}

\begin{figure*}
	\centering
	\includegraphics[clip=true,trim=0cm 0cm 0cm 0cm,width=0.24\textwidth]{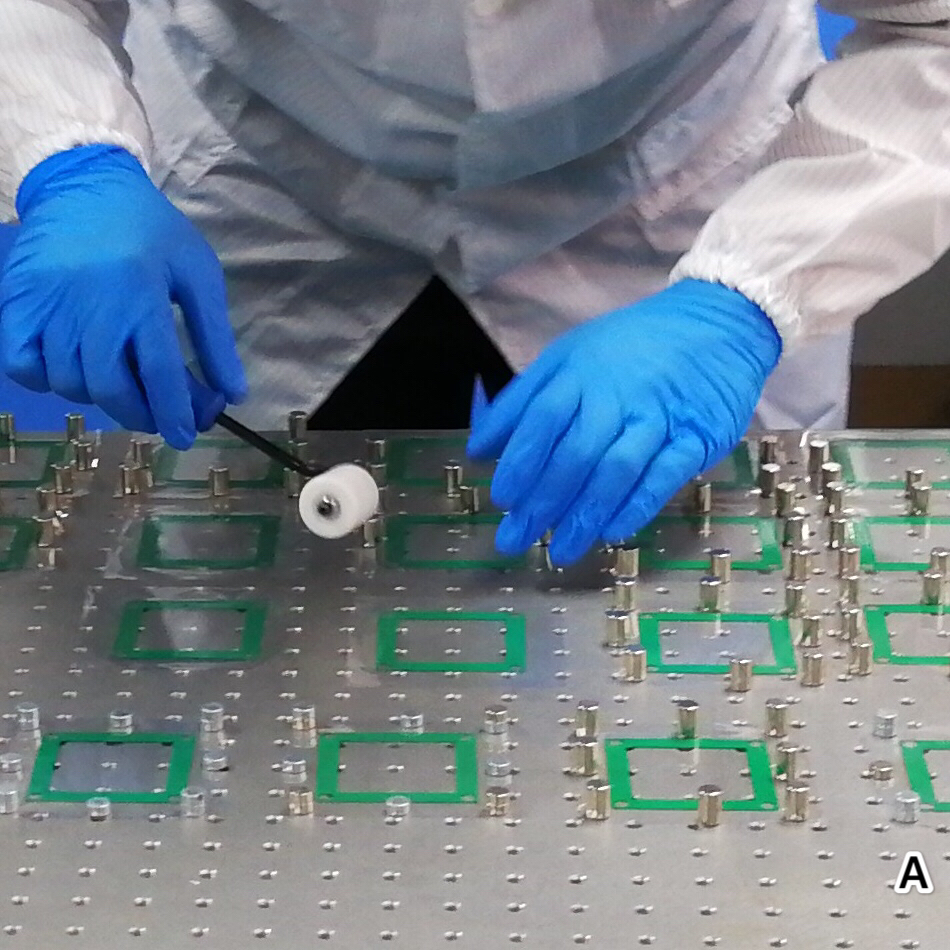}
	\includegraphics[clip=true,trim=0cm 0cm 0cm 0cm,width=0.24\textwidth]{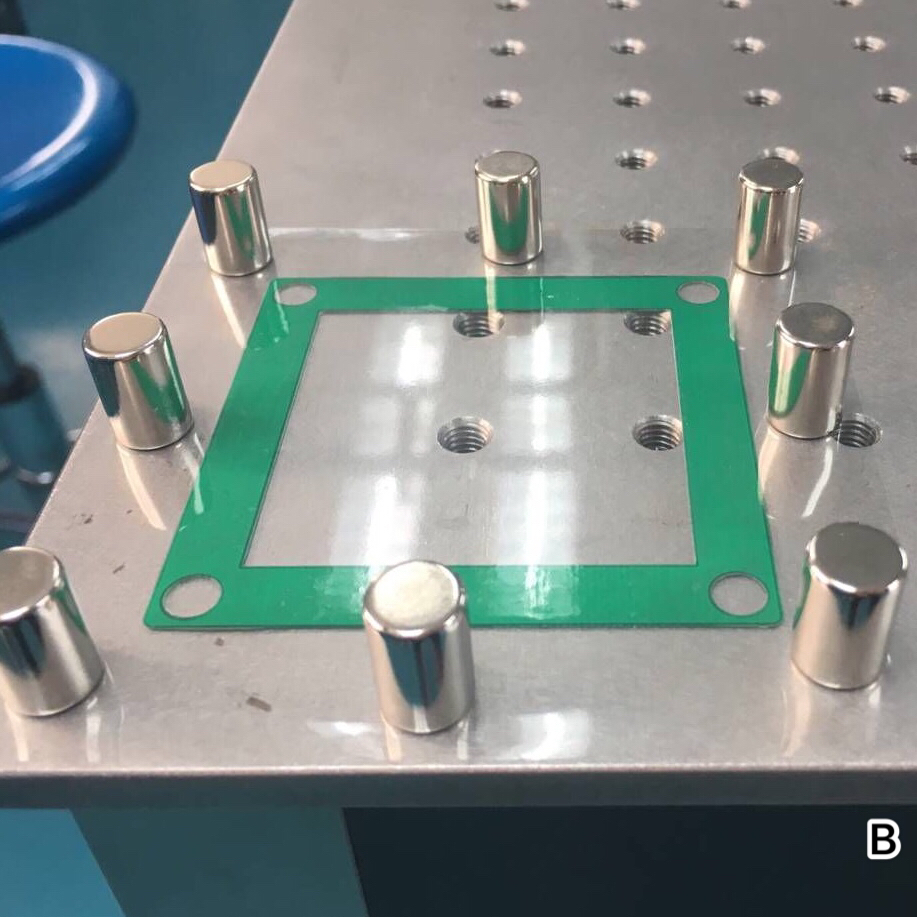}
	\includegraphics[clip=true,trim=0cm 0cm 0cm 0cm,width=0.24\textwidth]{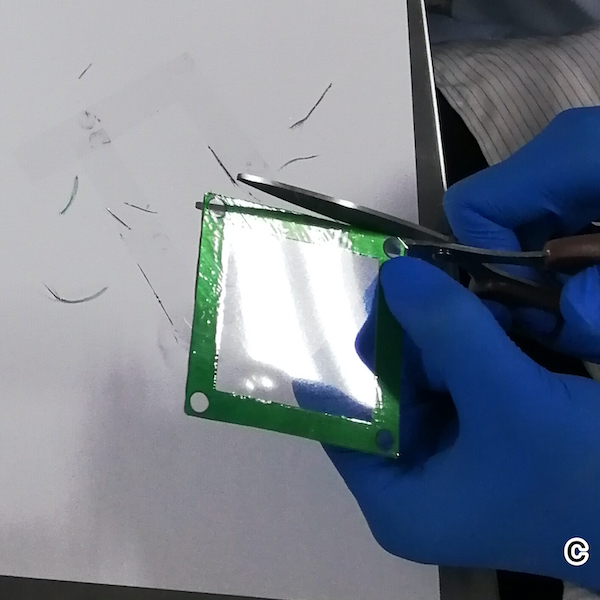}
	\includegraphics[clip=true,trim=0cm 0cm 0cm 0cm,width=0.24\textwidth]{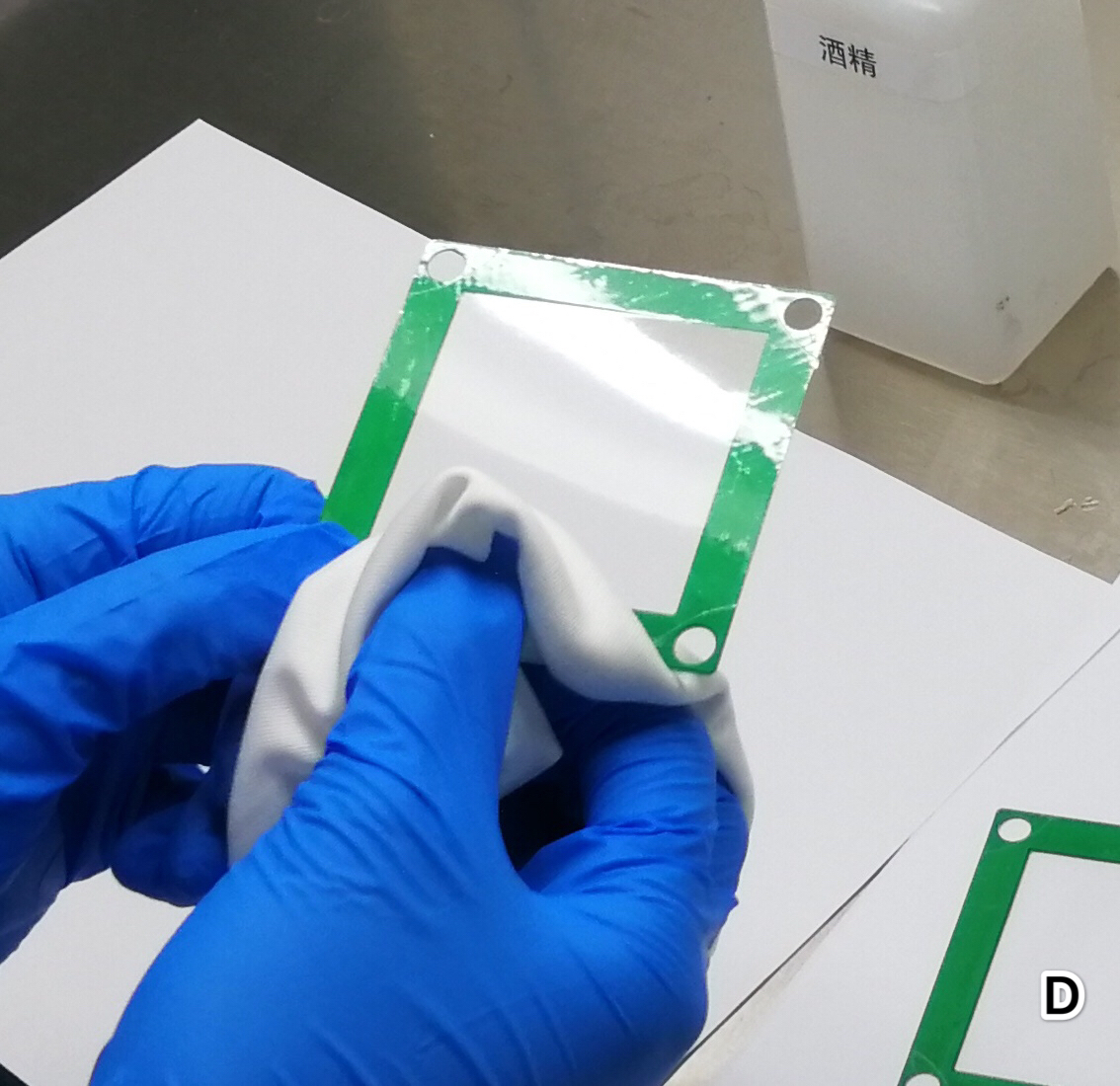}
	\includegraphics[clip=true,trim=0cm 0cm 0cm 0cm,width=0.24\textwidth]{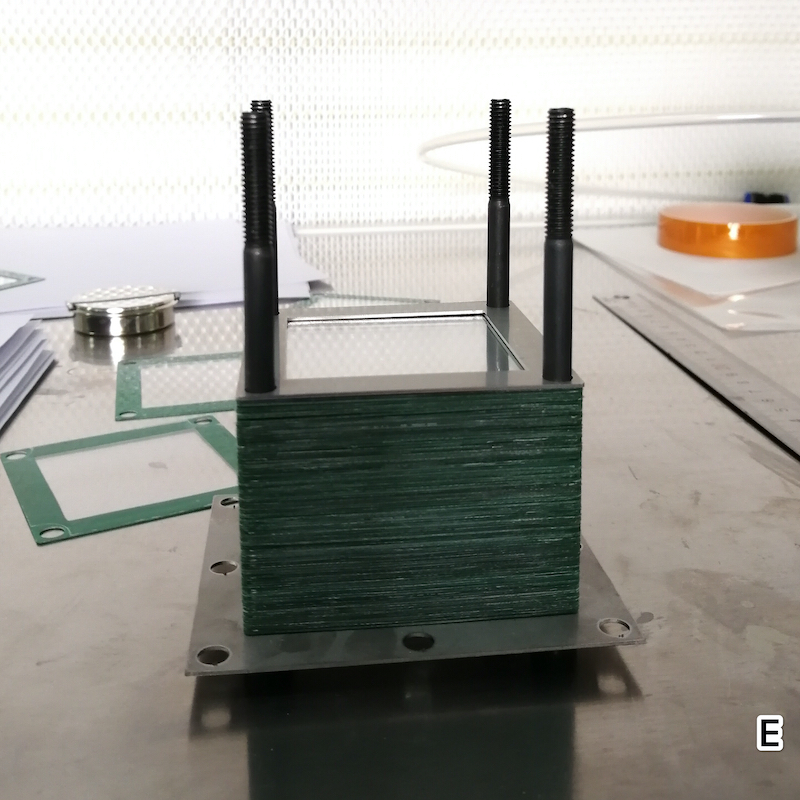}
	\includegraphics[clip=true,trim=0cm 0cm 0cm 0cm,width=0.24\textwidth]{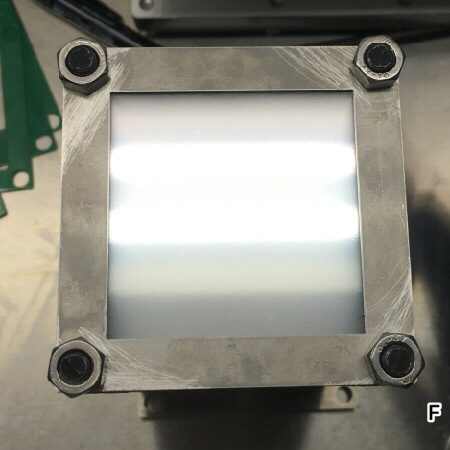}
	\includegraphics[clip=true,trim=0cm 0cm 0cm 0cm,width=0.24\textwidth]{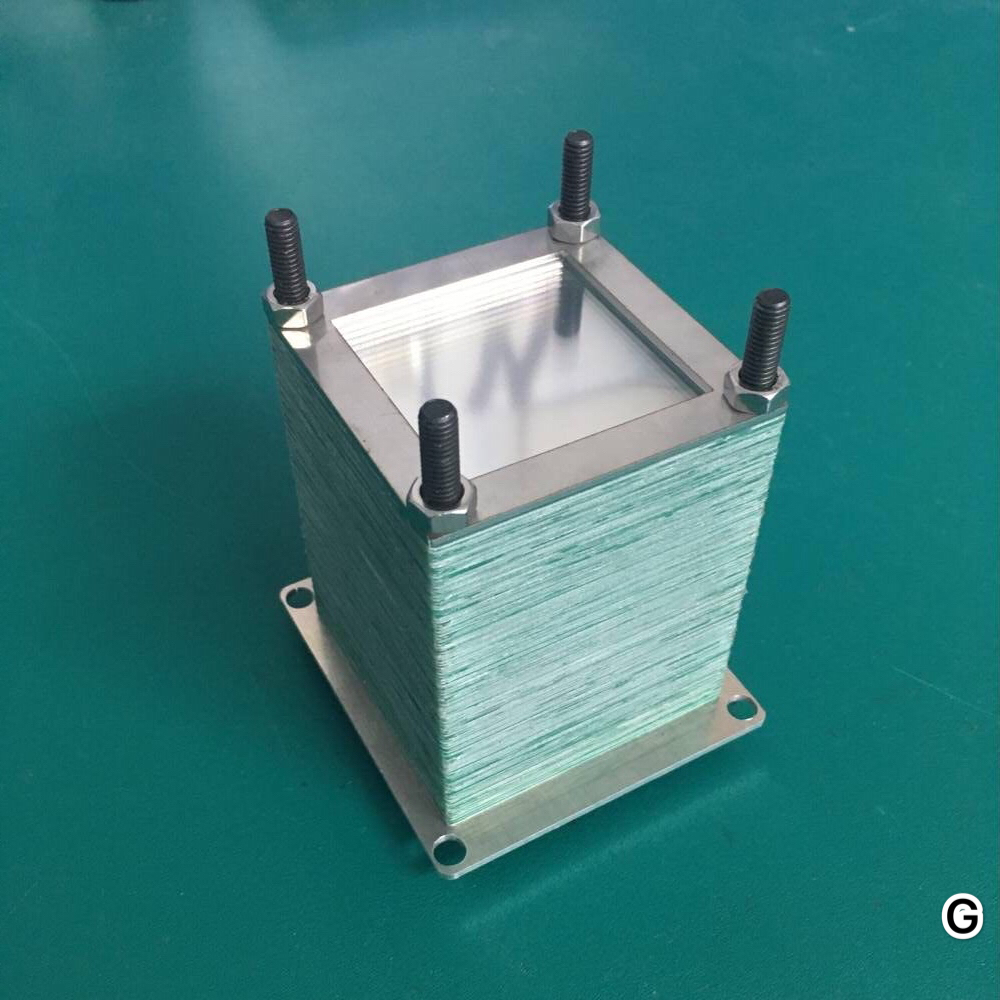}
	\includegraphics[clip=true,trim=0cm 0cm 0cm 0cm,width=0.24\textwidth]{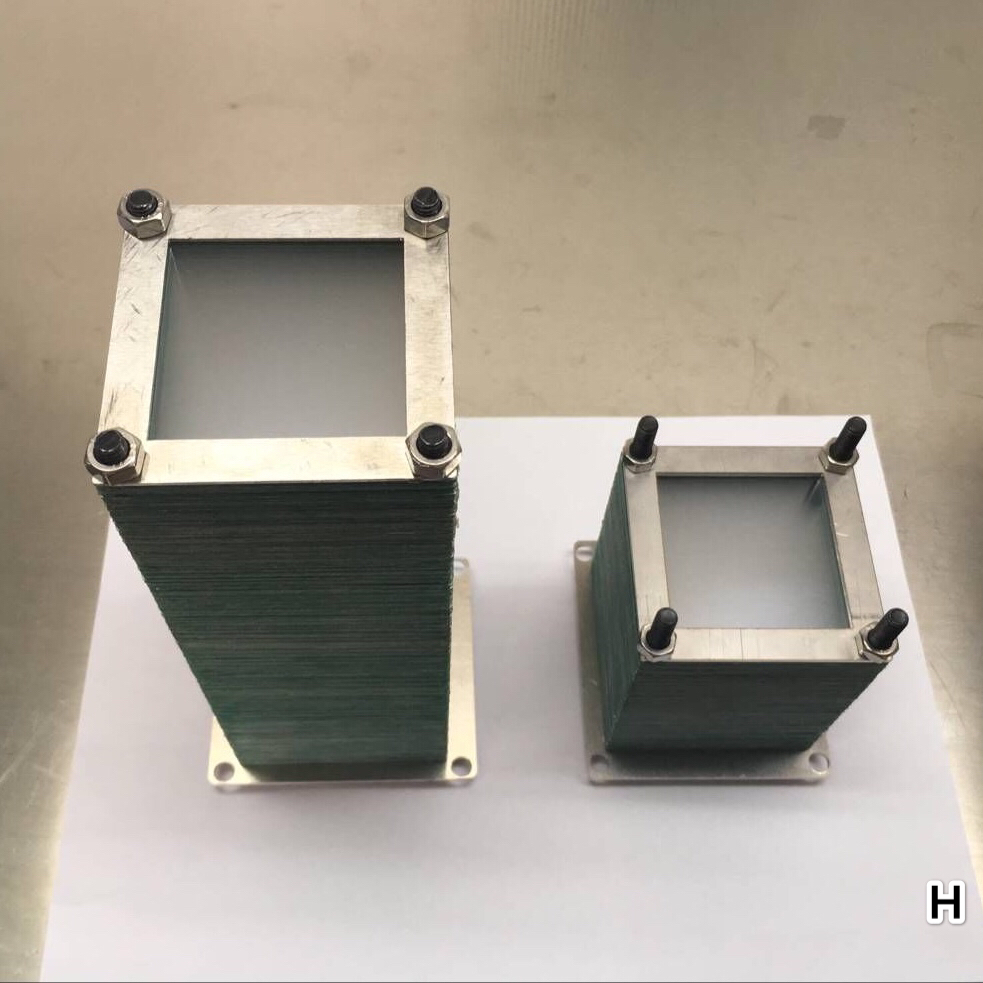}
	\caption{Different phases in the production of regular radiator prototypes.} 
	\label{fig:process}
\end{figure*}

According to the above simulation optimization procedure, we select a PP foil thickness of 20$\mu m$, and vary the air gap thickness and the number of foils to produce different radiators. In this process, critical aspects are to maintain the foil and gap thickness stable within acceptable limits, to make the foil surface smooth, and to use a light and thin supporting structure. We have manufactured regular radiator prototypes, using PP foils and PCB (Printed Circuit Board) spacer frames. The parameters of radiators are shown in Tab.\ref{tab:prototypes}, the different phases of production of the regular radiator prototypes are shown in Fig.\ref{fig:process} .

\begin{table}
	\centering
	\begin{tabular}{| c ||  c | c | c | c | }
		\hline
		Name & Material & $l_{1}(\mu m)$ & $l_{2}(\mu m)$ & $N$  \\
		\hline
		GXU0.5$\times$300 & PP+Air &  20  &  500  &  300  \\
		\hline
		GXU0.5$\times$225 & PP+Air &  20  &  500  &  225  \\
		\hline
		GXU0.5$\times$150 & PP+Air &  20  &  500  &  150  \\
		\hline
		GXU0.8$\times$300 & PP+Air &  20  &  800  &  300  \\
		\hline
		GXU0.8$\times$225 & PP+Air &  20  &  800  &  225  \\
		\hline
		GXU0.8$\times$150 & PP+Air &  20  &  800  &  150  \\
		\hline
	\end{tabular}
	\caption{Components and parameters of the regular radiator prototypes. }
	\label{tab:prototypes}
\end{table}

These can be summarized as follows:
\begin{enumerate}
	\item[A] We glue the PP foil and PCB spacer frame together with an epoxy resin adhesive. The air bubbles in the adhesive surface are removed with the roller. Then the PP foils were stretched to make their surface smooth and flat.
	\item[B] The epoxy resin adhesive is solidified after being fixed by eight magnets for 24 hours.
	\item[C] Once PP foil and PCB spacer frame (called PP-PCB) are glued together, the PP foil is cut away.
	\item[D] Then we rinse the foil surface with alcohol.
	\item[E] The different radiators were assembled by placing different numbers of PP-PCBs onto aluminum alloy bottom frames and fixing them by four screws. The final regular radiators are shown in F, G and H.
\end{enumerate}

Generally, both the material and the epoxy resin adhesive can be easily obtained. Thanks to this production process, each radiator interface is kept smooth and flat, and the foils are kept at the same distance.

\section{Transition radiation detector}
\label{sec:TRD}

The TR spectrum has a continuous shape and extends from the visible to the X-ray region, with the highest yield in X-ray range. Cherenkov radiation \citep{cherenkov1934visible}, which mainly concentrates in the visible and ultraviolet regions, may also be emitted when high-speed particles pass through the medium; thus the detector must be designed as an X-ray detector to reduce the background from Cherenkov radiation. The detector spectral response must match the X-ray frequency band of the radiator which has been optimized for maximal TR production. Since the direction of emission of the TR photons and the direction of the incident charged particles are almost collinear ($\theta\sim1/\gamma$), the TRD also detects the ionization energy loss of charged particles, which is the main background. 

A TRD converts TR photons into photoelectrons through the photoelectric effect \citep{kn2011}. Inert gases are commonly used in the detection. Xenon is a good medium for X-ray detection because of its large cross section for the photoelectric effect. Neon and Argon can also be used and have the advantage of low costs. The type of counting gas and its thickness affect the contribution from ionization energy loss induced by charged particles. The commonly used method of TR detection is to detect the sum of two signal: the ionization energy loss and TR.

\begin{figure*}
	\centering
	\includegraphics[clip=true,trim=0cm 0cm 0cm 0cm,width=0.255\textwidth]{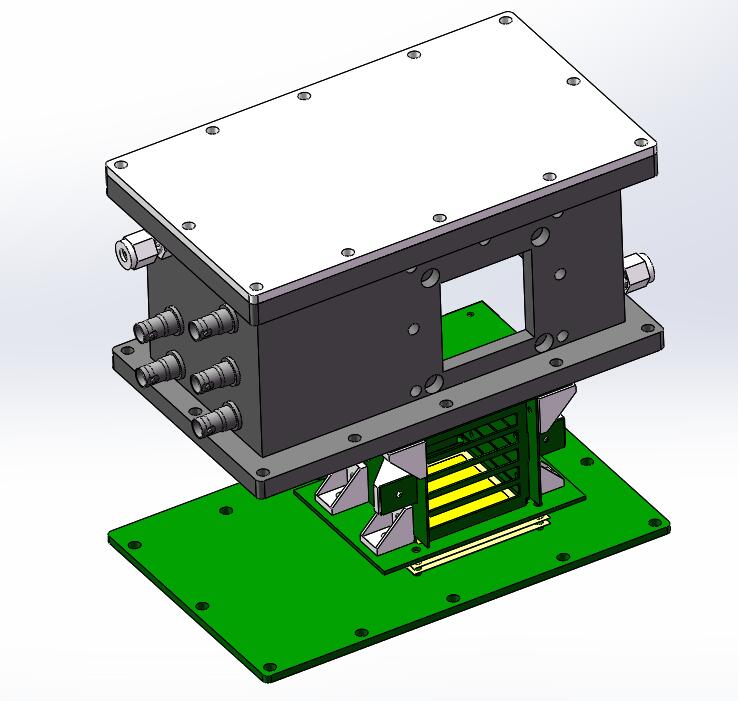}
	\includegraphics[clip=true,trim=0cm -1cm 0cm -1cm,width=0.32\textwidth]{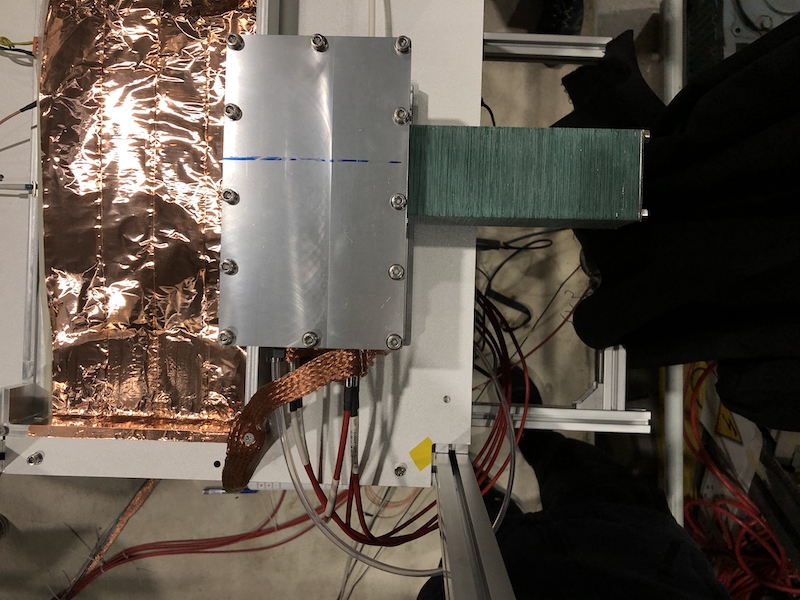}
	\includegraphics[clip=true,trim=0cm 0cm 0cm 0cm,width=0.32\textwidth]{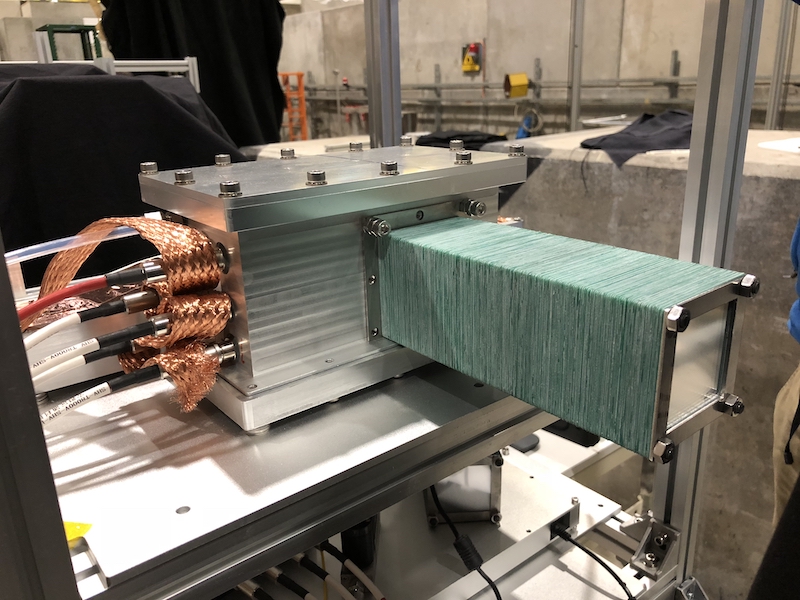}
	\caption{Scheme of the TRD (left) and its prototype (center and right).} 
	\label{fig:TRD}
\end{figure*}

The prototype of Side-On TRD which was tested at the CERN SPS is shown in Fig.\ref{fig:TRD}. As shown in the left picture, the chamber is made of aluminum alloy, and it is divided into two parts, body and cover. There are two side-on windows of the same size symmetrically placed on the lateral walls of the chamber, one of which is used to fix the radiator. On the other side, there are five high-voltage connectors, which provide high voltage for the Thick Gaseous Electron Multiplier (THGEM) \citep{Breskin:2008cb} and the field cage. Two holes are used to provide counting gas for the detector and to keep the gas composition constant inside the chamber.

\begin{figure}
	\centering
	\includegraphics[clip=true,trim=1cm 1cm 1cm 1cm,width=0.5\textwidth]{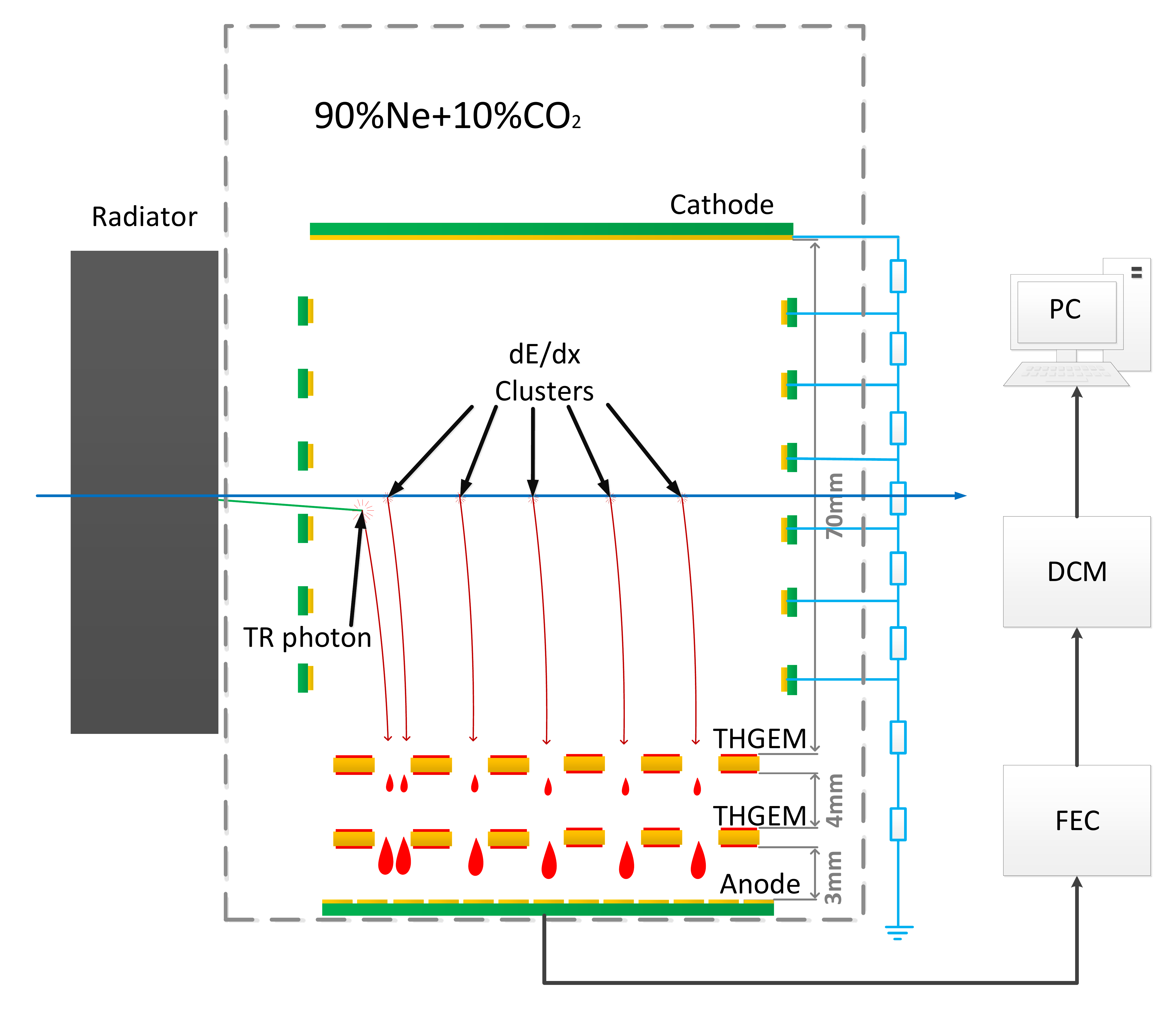}
	\caption{The schematic diagram of the Side-On TRD. The grey object on the left is the regular radiator, the blue line indicates the track of charged particles, the green line indicates the track of TR photons, red lines indicate the electron tracks, red droplets represent the electron clusters. The anode readout system is seqmented into 64 channels with the total width of 5$cm$. The Front-End Card (FED) and Data Collection Module (DCM) are part of electronic readout system.} 
	\label{fig:str}
\end{figure}

The detector thickness has to be chosen carefully to absorb a maximal amount of TR photons and also minimize the ionization energy loss of the charged particles. Actually, the peak of the TR spectrum generated by GXU0.5$\times$300 mainly concentrates in the interval 6-14 keV. In order to reduce the background from energy loss, the thickness has to be kept low enough, but that would also decrease the detection efficiency. Here we design a new prototype of TRD with 64 channels of readout electronics, called Side-On TRD, to reduce this problem. A scheme of the side-on TRD filled with 5-cm counting gas is shown in Fig.\ref{fig:str}. When charged particles pass through the radiator and enter into the detector, secondary electrons produced along the track are drifted vertically towards the anode by the applied electric field. If a TR photon is produced, it will be converted into a photoelectron by photoelectric effect in the gas; the electron will drift vertically to the corresponding channel of the anode. The signal in channels detecting photoelectrons from TR absorption is on average stronger than the signal from ionization energy loss. In this way channels with TR signal can to some extend be separated from the background due to energy loss.

\section{Results of test beam experiment}
\label{sec:EC}

\begin{figure}
	\centering
	\includegraphics[clip=true,trim=2cm 0cm 0cm 0cm,width=0.58\textwidth]{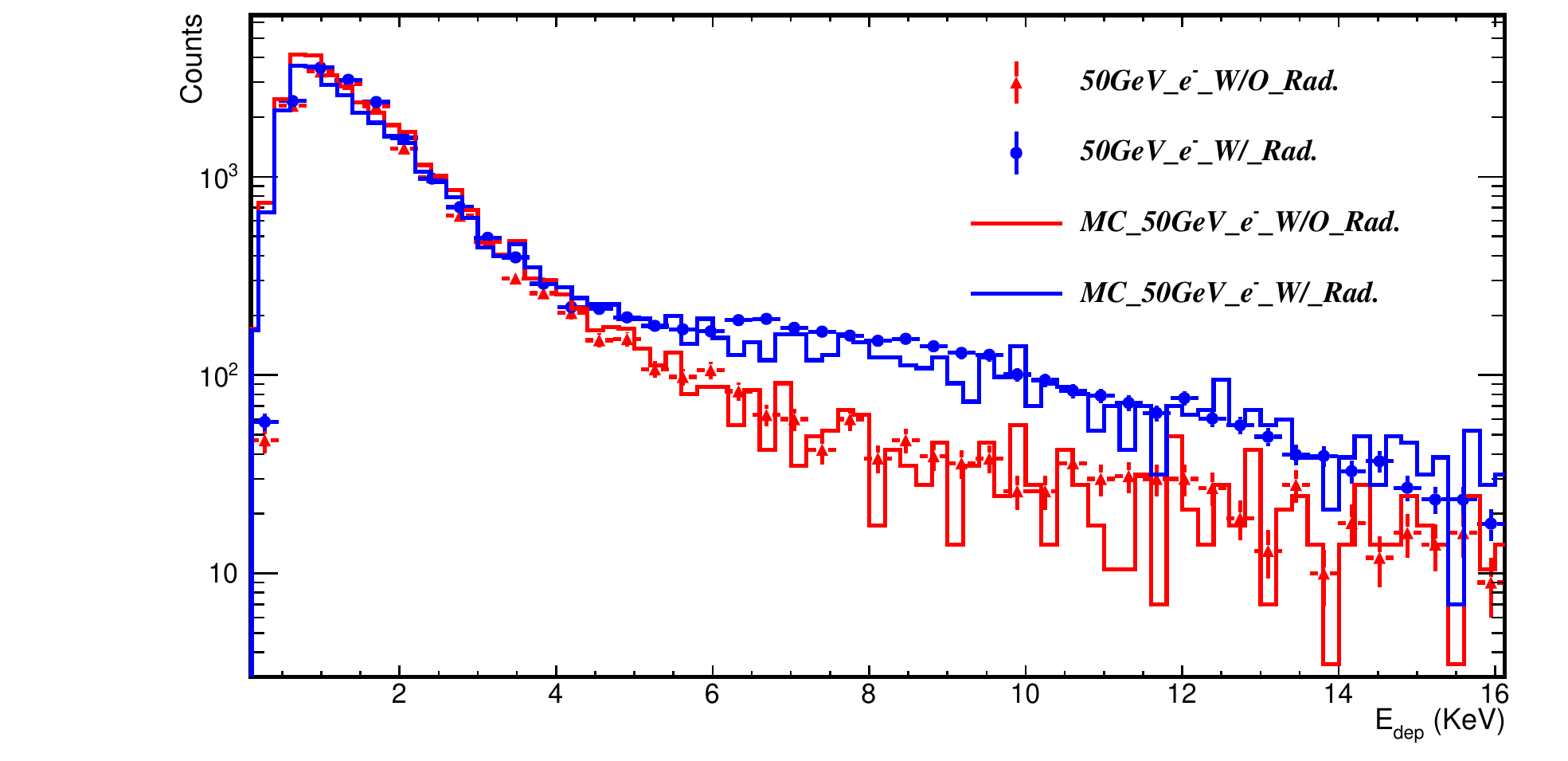}
	\caption{Energy spectra registered in the Side-On TRD for a 50 GeV electron beam with and without radiators. The points of different style indicate the data from the CERN SPS test beam experiment, and the lines are from simulation. Blue: with radiator; Red: without radiator. The radiator is GXU0.5$\times$300, the counting gas in the detector is a mixture of 90\%Ne+10\%CO$_{2}$.} 
	\label{fig:significance}
\end{figure}

\begin{figure*}
	\centering
	\includegraphics[clip=true,trim=2cm 0cm 0cm 0cm,width=0.58\textwidth]{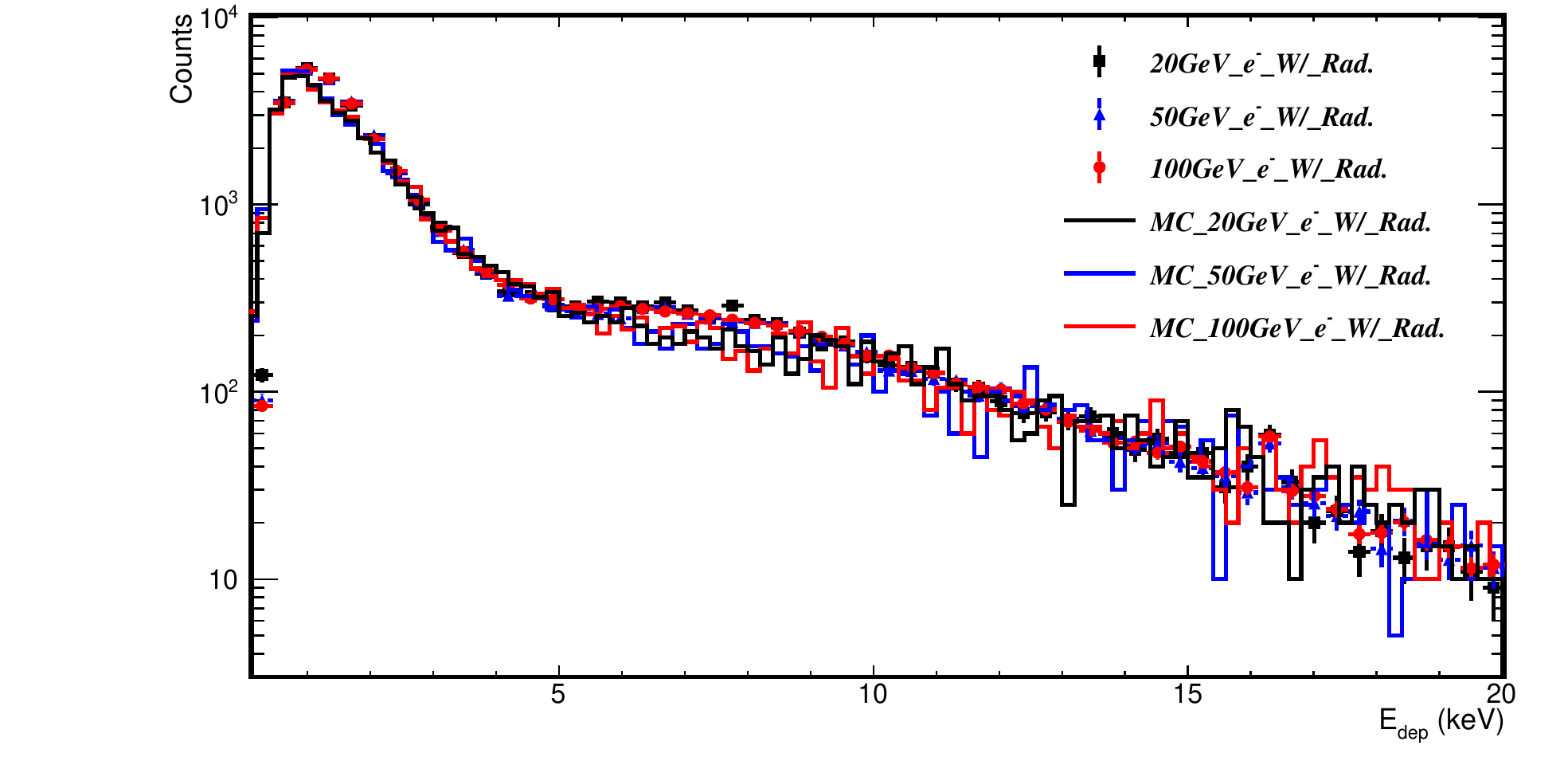}
	\caption{Energy spectra registered in the Side-On TRD with radiator for electron beams of different energies. The points of different style indicate the data from the CERN SPS test beam experiment, and the lines are from simulation. Black: 20 GeV electron beams; Blue: 50 GeV electron beams; Red: 100 GeV electron beams. The radiator is GXU0.5$\times$300, the counting gas in the detector is a mixture of 90\%Ne+10\%CO$_{2}$.} 
	\label{fig:spec}
\end{figure*}

To verify the performance of the Side-On TRD prototype and its readout system, test beam data were taken at the CERN SPS in November 2018. We use the TRD with and without radiator to start test beam experiment. The radiator and counting gas are separated by PP foil with the thickness of 100 $\mu m$. Since the electric field is not be evenly distributed at the edges of the field cage, we select only experimental data from the middle channels between 12-53 to minimize the effect. The same experimental conditions and selection of the channels were applied in the simulation. In the simulation setup, the regular radiator consists of 20 $\mu m$ PP foil and 800 $\mu m$ air gap, and the number of layers is 300; the X-ray detector is fulled of a mixture gas of 90\%Ne+10\%CO$_{2}$; the type of the incident particle beam is electron, with the energy of 20, 50 and 100 GeV. The detector is divided into 64 parts, and recorded the energy deposition of electron and TR, which is equivalent to 64-channel readout system in electronics. A comparison of the simulation results and experimental data was done for particle energies of 20, 50, 100GeV. As an example, spectra averaged over all channels which detected TR signals for a 50 GeV electron beam are shown in Fig. \ref{fig:significance}. We can see that experimental data and simulation are in good agreement. Despite the use of neon as counting gas, which has very low efficiency for TR photons, the TR signal recorded in the Side-On TRD is significantly above the signal from the background of ionization energy loss. The spectra for different energies of the electron beam are shown in Fig. \ref{fig:spec}. It is not surprising that the spectra essentially do not change since TR is already saturated at the available electron beam energies.

\section{Discussion and Conclusion}
\label{sec:CON}

The measurements with the Side-On TRD at the CERN SPS were carried out at atmospheric gas pressure, under a continuous flow of Ne/CO$_{2}$ (90:10). For the purpose of application in space the TRD should be sealed. For this reason, we are going to improve the air-tightness of the Side-On TRD in future developments. Moreover, we will test the Side-On TRD with different gases, different beam energies and particles, and other influencing factors, to better evaluate the detector performance.

A Side-On TRD simulation model was constructed using Geant4 according to the structure of the detector. The Monte Carlo simulation is consistent with the test beam results, which provides a good method for us to better understand the Side-On TRD. Moreover, complete simulation including electronics should be done in further works. Further study is needed for the calibration of the calorimeter in the energy range between 1 and 10 TeV by the Side-On TRD using Monte Carlo simulation.

There are systematical errors in the experimental process, such as the test beam instability, which would affect the experimental results and could be reduced by proper experimental design. The statistical error could be reduced by increasing the experimental statistics. As to the further work concerning the energy calibration, the expected error mainly comes from statistics and the uncertainty of the TRD response curve. We will increase statistics (i.e. increase the time of TRD in orbit) and optimize the Side-On TRD to reduce the errors.

In this work, we propose a new prototype of TRD called Side-On TRD to detect TR signals. We mainly focus on the following aspects in this paper:
\begin{enumerate} 
	\item Through optimization of both, radiators and TRD, we design a new prototype aimed at increasing the radiation yield, and we develop a new TRD prototype called Side-on TRD to match the TR photon frequency band efficiently.
	\item The Side-On TRD uses THGEM foils for gas amplification and a 64 channels electronic readout system, which is designed improve the TR signal detection.
	\item We use the optimized regular radiator and Side-On TRD in a test beam experiment at the CERN SPS. The experimental results are consistent with the simulation.
\end{enumerate}

\section*{Acknowledgments}

This work would not have been possible without the support of teachers and schoolmates in Guangxi University. The author would also like to thank the HERD collaborators for their efforts to make the test beam a success and acknowledge the useful discussions with Qian Liu from the University of Chinese Academy of Sciences, Changqing Feng from the University of Science and Technology of China.  The project was supported by Department of Physics and GXU-NAOC Center for Astrophysics and Space Sciences, Guangxi University. This work is supported by the National Natural Science Foundation of China
(grant Nos. U1731239, 11773007, 11533003, 11851304), the Guangxi Science Foundation (grant Nos. 2018GXNSFGA281007, 2018GXNSFFA281010, 2018GXNSFDA281033, 2017AD22006, 2018GXNSFGA281005), the Innovation Team and Outstanding Scholar Program in Guangxi Colleges, the One-Hundred-Talents Program of Guangxi colleges.

\begin{appendix} 
	
\section{TR Calculation method for regular radiators}
\label{app:TR}
	
\subsection{Single foil}\label{subsec:single}
	
In the case of $\gamma\gg1, \xi^2=\omega_{P}^{2}/\omega^{2}\ll1, \theta\ll1$ ($\theta$ is the emission angle of TR photons with respect to the direction of the charged particle), the transition radiation intensity of a single interface can be written as 
\begin{equation}\label{eqA1}
\biggl(\frac{d^{2}W}{d\omega d\theta} \biggr) _{interface}=
\frac{2\alpha\theta^{3}}{\pi}\biggl(\frac{1}{\gamma^{-2}+\theta^{2}+\xi_{1}^{2}}-\frac{1}{\gamma^{-2}+\theta^{2}+\xi_{2}^{2}}\biggr)^{2}
\end{equation}
where $\gamma$ is the Lorentz factor of the incident charged particle, $\alpha=e^{2}/\hbar c\approx 1/137$ is the fine structure constant. $\omega_{P,i}$ is the plasma frequency of the medium, represented by
\begin{equation}\label{eqA2}
\omega_{P}=\sqrt{\frac{4\pi \alpha n_{e}}{m_{e}}}\approx28.8\sqrt{\frac{\rho Z}{A}} eV
\end{equation}
here $m_{e}, N_{A}, A, \rho$ are electron mass, Avogadro constant, average relative atomic mass and medium density, respectively \citep{aj2001}. The average electron density $n_{e}$ is expressed as
\begin{equation}\label{eqA3}
n_{e}=\frac{\rho N_{A} Z}{A}
\end{equation}
For the compound or mixture, $Z$ and $A$ can be approximated by using the relative weighting factor $w_{i}$:
\begin{equation}\label{eqA4}
Z=\sum_{i}{w_{i} Z_{i}},  A=\sum_{i}{w_{i} A_{i}}
\end{equation}
For example, the plasma frequencies of polypropylene and air are $\omega_{P}^{CH_{2}} \approx 20.65 eV$ and $\omega_{P}^{Air} \approx 0.71 eV$, respectively.
	
In practice, Eq.\ref{eqA1} is the basic for the transition radiation. It can be integrated over $\theta$ to obtain the differential energy spectrum:
\begin{equation}\label{eqA5}
\biggl(\frac{dW}{d\omega}\biggr)_{interface}=\frac{\alpha}{\pi}\biggl[\biggl(\frac{\omega_{P,1}^{2}+\omega_{P,2}^{2}+2\omega /\gamma^{2}}{\omega_{P,1}^{2}-\omega_{P,2}^{2}}\biggr)
\times\ln\biggl(\frac{\gamma^{-2}+\omega_{P,1}^{2}/\omega^{2}}{\gamma^{-2}+\omega_{P,2}^{2}/\omega^{2}}\biggr)-2\biggr]
\end{equation}
	
\begin{figure*}[ht]
\centering
\includegraphics[clip=true,trim=3cm 6cm 4cm 6cm,width=0.45\textwidth]{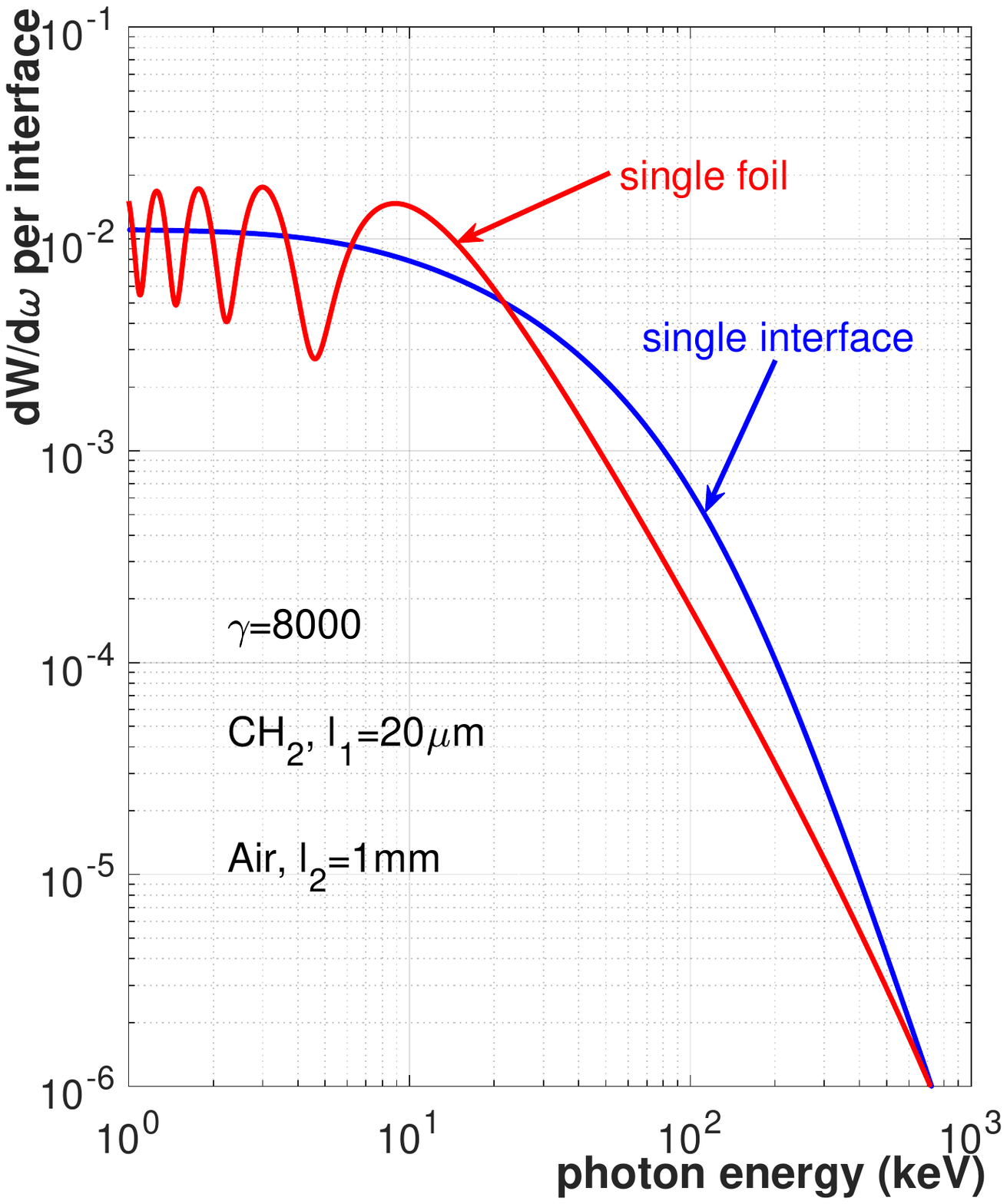}
\includegraphics[clip=true,trim=3cm 6cm 4cm 6cm,width=0.45\textwidth]{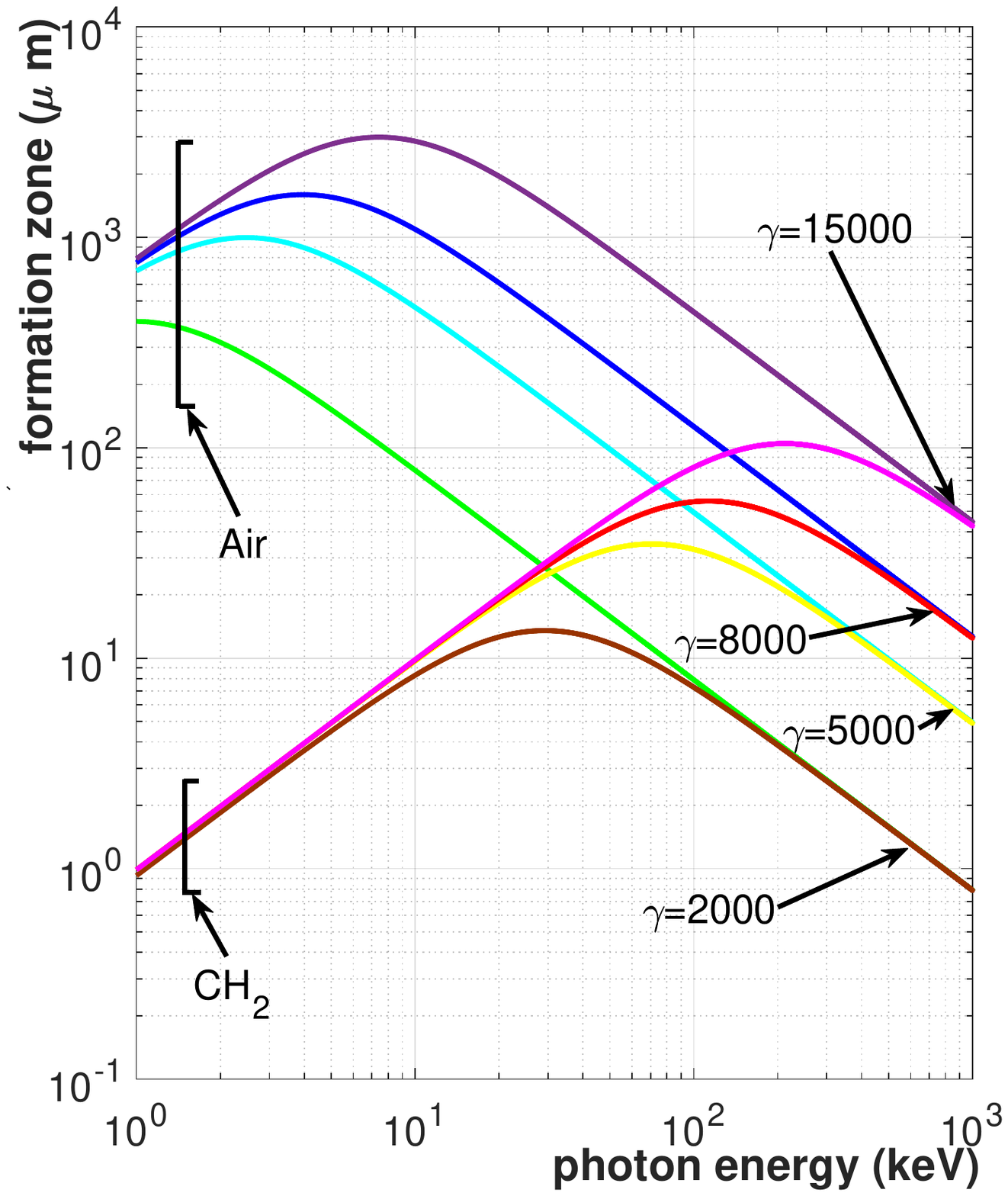}
\caption{Left: TR-spectrum of single foil (red line) and single interface (blue line). Right: formation zone of air and CH$_{2}$ as a function of the photon energy for different Lorentz factor.} 
\label{fig:SPandFMZ}
\end{figure*}

For the single foil (two interfaces), it needs to sum up the contributions from both interfaces of the foil to the surrounding medium, then it can be described by the following expression:
\begin{equation}\label{eqA6}
\biggl(\frac{d^{2}W}{d\omega d\theta}\biggr)_{singlefoil}=\biggl(\frac{d^{2}W}{d\omega d\theta}\biggr)_{interface}\times 4\sin^{2}{\frac{\varphi_{1}}{2}}
\end{equation}
where $4\sin^{2}{\varphi_{1}/2}$ is the interference factor, $\varphi_{i}\approx {l_{i}}/{D_{i}}$; the average amplitude modulation $\langle 4\sin^{2}{\varphi_{1}/2}\rangle _{\bigtriangleup \omega,\bigtriangleup \theta}\approx 2$ when $l_{1}\gg D_{1}$ as discussed in Ref. \citep{mlc1974}. TR spectra for single interface and single foil with the same medium are shown in
Fig.\ref{fig:SPandFMZ} (left). The medium thickness is $l_{i}$, and the phase $\varphi_{i}$ can be expressed as:
\begin{equation}\label{eqA7}
\varphi_{i}=(\gamma^{-2}+\theta^{2}+\xi_{i}^{2})(\omega l_{i}/2\beta c)\approx l_{i}/D_{i},(\beta\approx 1)
\end{equation}
$D_{i}$ is defined as the formation zone of a medium, and it can be expressed as
\begin{equation}\label{eqA8}
D_{i}=\frac{2c}{\omega}(\gamma^{-2}+\theta^{2}+\xi_{2}^{2})^{-1}
\end{equation}
	
Physically, the formation zone is the distance along the particle trajectory in a given medium after which the separation between particle and the generated photon is of the order of the photon wavelength \citep{mlc1974}. The formation zone depends on the Lorentz factor $\gamma$ of the charged particle, the TR photon energy and the plasma frequency of the medium. The formation zone (for air and polypropylene) dependence on the photon energy is shown in Fig.\ref{fig:SPandFMZ} (right) for different values of the Lorentz factor of the charged particle.
	
Due to this dependence, it can be noticed from Eq. \ref{eqA7} that, when $l_{1}>>D_{1}$, the phase angle $\varphi_{1}$ changes rapidly with the energy of the photon; in this case the average radiation energy of each layer is similar to that of a single interface. When instead $l_{1}<<D_{1}$, the phase angle $\varphi_{1}$ is very small, and the average radiation is suppressed by destructive interference.
	
\subsection{Multi-layer}\label{subsec:multi}
	
Transition radiation from a stack of $N$ regular foils of thickness $l_{1}$ separated by distances $l_{2}$, can be described by the following expression:
\begin{equation}\label{eqA9}
\biggl(\frac{d^{2}W}{d\omega d\theta}\biggr)_{stack}=\biggl(\frac{d^{2}W}{d\omega d\theta}\biggr)_{singlefoil}\times\frac{\sin^{2}[N(l_{1}/D_{1}+l_{2}/D_{2})]}{\sin^{2}(l_{1}/D_{1}+l_{2}/D_{2})}
\end{equation}
According to Refs. \citep{mlc1974}, when $l_{2}>D_{2}$, the last term in Eq.\ref{eqA9} can be replaced by its average
\begin{equation}\label{eqA10}
\biggl\langle\frac{\sin^{2}[N(l_{1}/D_{1}+l_{2}/D_{2})]}{\sin^{2}(l_{1}/D_{1}+l_{2}/D_{2})}\biggr\rangle_{\bigtriangleup \omega, \bigtriangleup \theta}\approx N
\end{equation}
Thus, the observed yield in this case is well approximated by $N$ times the radiation yield of a single foil. An approximate analytical integration of Eq.\ref{eqA9} over TR emission angles $\theta$ gives:
\begin{equation}\label{eqA11}
\biggl(\frac{dW_{N}}{d\omega}\biggr)_{stack}=2\alpha\hbar cN(\omega_{P,1}^{2}-\omega_{P,2}^{2})\frac{(l_{1}+l_{2})^{2}}{\omega^{2}}\times\sum_{r_{min}}^{r_{max}}\frac{(r-R_{m})\sin^{2}[\pi(r-R_{1})l_{2}/(l_{1}+l_{2})]}{(r-R_{1})^{2}(r+R_{2})^{2}}
\end{equation}
where
	
$R_{1,2}=el_{1,2}(\omega_{P,1}^{2}-\omega_{P,2}^{2})/4\pi\hbar c\omega$,
	
$R_{m}=\frac{e}{4\pi\hbar c}\biggl[\frac{\omega(l_{1}+l_{2})}{\gamma^{2}}+\frac{l_{1}\omega_{P,1}^{2}+l_{2}\omega_{P,2}^{2}}{\omega}\biggr]$,
	
$r_{min}=\frac{l_{1}+l_{2}}{2\pi\gamma c}\biggl(\frac{l_{1}\omega_{P,1}^{2}+l_{2}\omega_{P,2}^{2}}{l_{1}+l_{2}}\biggr)^{1/2}$,
	
$r_{max}=\gamma r_{min}$.
	
Taking into account the losses by absorption effect of the foils and gaps, the average value of Eq.\ref{eqA9} becomes:
\begin{equation}\label{eqA12}
\biggl(\frac{d^{2}W}{d\omega d\theta}\biggr)_{foil}=\biggl(\frac{d^{2}W}{d\omega d\theta}\biggr)_{stack}\times\exp\biggl(\frac{\sigma(1-N)}{2}\biggr)\frac{\sin^{2}(N\phi_{12}/2)+\sinh^{2}(N\sigma/4)}{\sin^{2}(\phi_{12}/2)+\sinh^{2}(\sigma/4)}
\end{equation}
where $\phi_{12}=\phi_{1}+\phi_{2}$ is the phase difference and $\phi_{i}=(\gamma^{-2}+\theta^{2}+\xi_{i}^{2})\omega l_{i}/2$, the absorption cross section of the radiator materials is $\sigma=\sigma_{1}+\sigma_{2}=\mu_{1}l_{1}+\mu_{2}l_{2}$ (foil + gap), $\mu_{1,2}$ being the absorption coefficient of medium \citep{aj2001}. 
	
\subsection{Main characteristics}
The TR basic characteristics can be summarized as follows
\begin{enumerate}
\item When a charged particle passes through the boundary between two different media with plasma frequencies $\omega_{P,1}$ and $\omega_{P,2}$, the TR energy is:
\begin{equation}\label{eqA13}
W=\frac{\alpha\hbar\gamma(\omega_{P,1}-\omega_{P,2})^{2}}{3(\omega_{P,1}+\omega_{P,2})}
\end{equation}
Thus it can be seen that the total energy of the TR photons is proportional to the Lorentz factor of the incident particle $\gamma$. This property can be exploited for high energy particle calibration.
\item The emission angle $\theta$ between the photons and the moving charged particles is inversely proportional to $\gamma  (\theta\sim1/\gamma)$. For the condition of $10^{3}<\gamma<10^{4}$, TR is emitted forward.
\item TR-spectrum is a continuous spectrum. It extends from the visible to the X-ray region, but mainly concentrates in the X-ray energy region.
\item The TR spectrum shows a peak at
\begin{equation}\label{eqA14}
\omega_{max}=\frac{l_{1}\omega_{P,1}^{2}}{2\pi\beta c}
\end{equation}
By varying the material and thickness of the radiator foil the TR-spectrum peak can be adjusted at the proper energy position.
\item For the multi-layer radiators, in order to produce more photons, the interference conditions $l_{1}\gg D_{1}$ and $l_{2}>D_{2}$ has to be satisfied. The average TR intensity is approximately $N$ times that of a single foil.
\end{enumerate}
	
\section{The properties and attenuation coefficients of several common foil materials and gap materials}
\label{app:xcom}
	
\begin{figure*}
	\centering
	\includegraphics[clip=true,trim=0cm 7cm 0cm 8cm,width=0.48\textwidth]{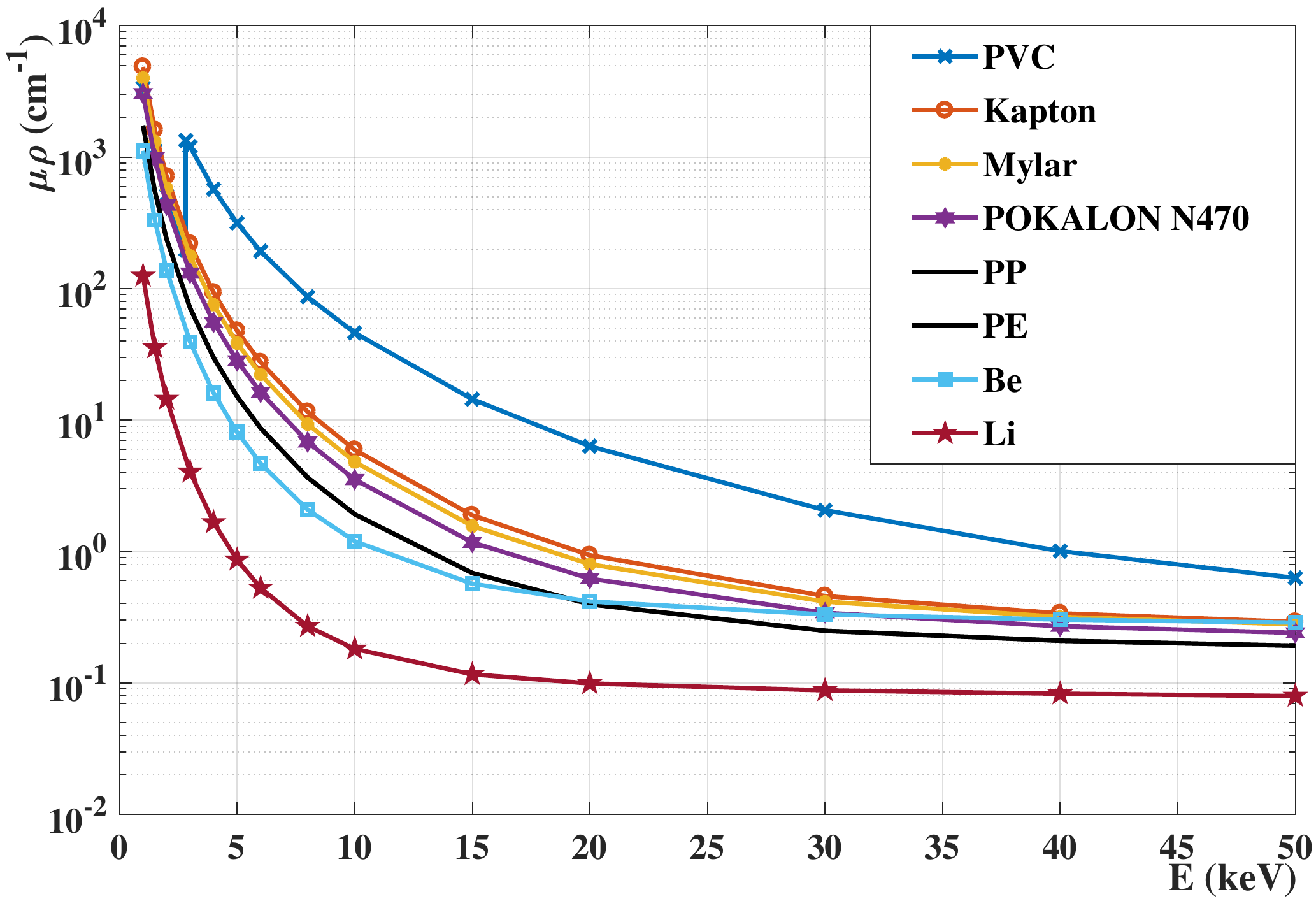}
	\includegraphics[clip=true,trim=0cm 7cm 0cm 8cm,width=0.48\textwidth]{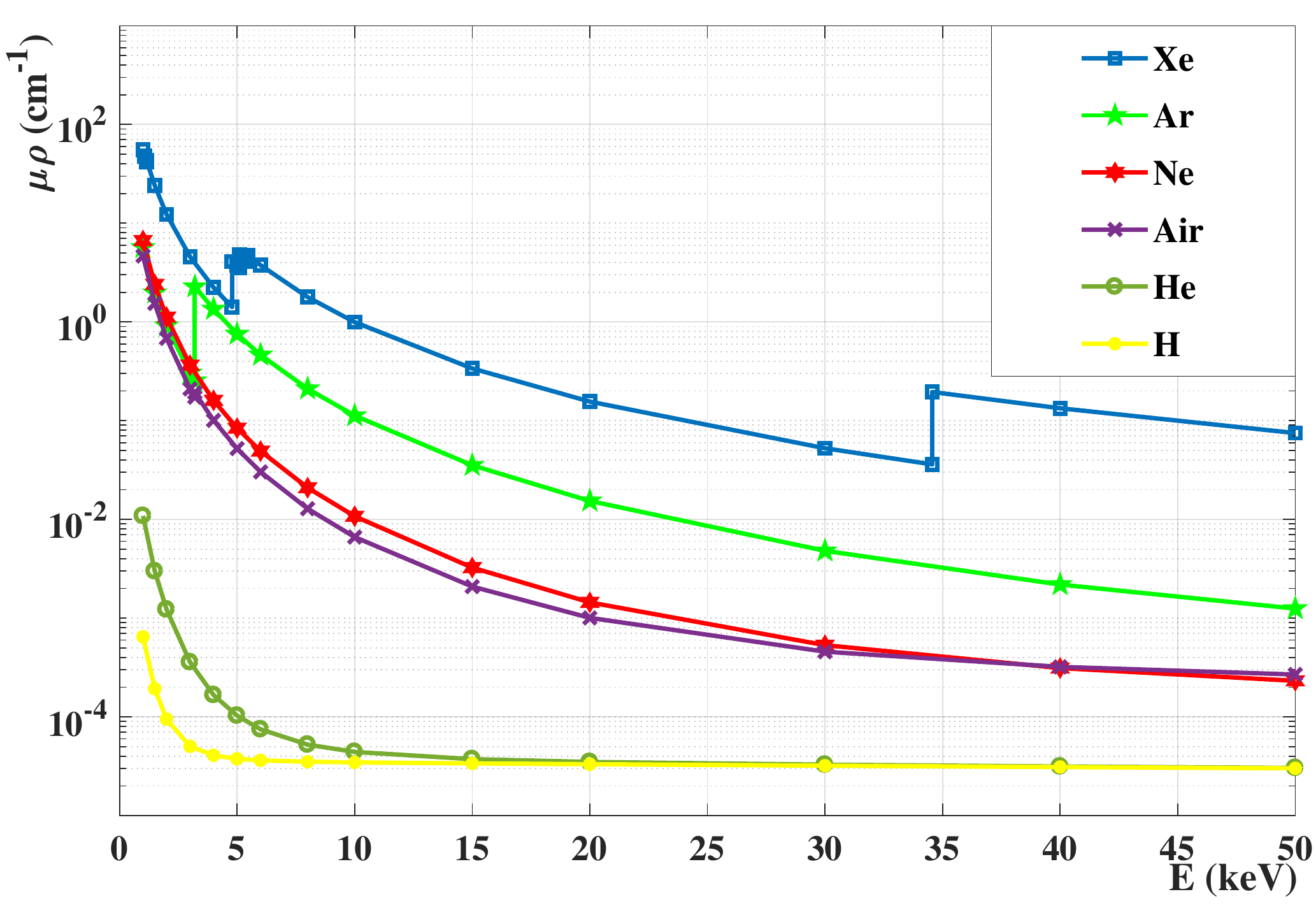}
	\caption{Attenuation coefficients $\mu$ times density $\rho$ as a function of photon energy $E$ for different foil (left) and gap (right) materials. $\mu$ is obtained by Photon Cross Sections Database \citep{jb1998}. In the left-hand plot, multiplication sign: PVC; hollow circle: Kapton; solid circle: Mylar; hexagram: POKALON N470; solid line: PP and PE; square: Be; pentagram: Li. In the right-hand plot, squre: Xe; pentagram: Ar; hexagram: Ne; multiplication sign: air; hollow circle: He; solid circle: H.} 
	\label{fig:Absor}
\end{figure*}

The material properties, which are essential for this optimization, are shown in tab. \ref{tab:material} and Fig. \ref{fig:Absor}.
	
\begin{table*}
\centering
		
\begin{tabular}{| c ||  c | c | c | c | c | }
	\hline
	Foil material & Chemical formula & $Z$ & $A$ & $\rho$(g/cm$^{3}$) &  $\omega_P$(eV) \\
	\hline
	Mylar & $(C_{10}H_{8}O_{4})_{n}$ &  4.55  &  8.73  &  1.393  &  24.53  \\
	\hline
	Polyethylene(PE)& $(C_{2}H_{4})_{n}$ &  2.67  &  4.67  &  0.900  &  20.65  \\
	\hline
	Polypropylene(PP) & $(C_{3}H_{6})_{n}$ &  2.67  &  4.67  &  0.900  &  20.65  \\
	\hline
	Polyvenylchlorid(PVC) & $(C_{2}H_{3}Cl)_{n}$ &  6.33  &  10.33  &  1.300  &  25.02  \\
	\hline
	Kapton & $(C_{22}H_{10}N_{2}O_{5})_{n}$ &  5.03  &  9.79  &  1.430  &  24.67  \\
	\hline
	Lithium & $Li$ &  3.00  &  6.94  &  0.534  &  13.84  \\
	\hline
	Beryllium & $Be$ &  4.00  &  9.01  &  1.848  &  26.09  \\
	\hline
	POKALON N470 & $(C_{16}H_{14}O_{3})_{n}$ &  4.06  &  7.70  &  1.150  &  22.43  \\
	\hline
	\hline
	Gap material & Chemical formula & $Z$ & $A$ & $\rho$(g/cm$^{3}$) &  $\omega_P$(eV) \\
	\hline
	Hydrogen & $H$ &  1.00  &  1.01  &  8.38e-05  &  0.26  \\
	\hline
	Helium & $He$ &  2.00  &  4.00  &  1.70e-04  &  0.27  \\
	\hline
	Neon & $Ne$ &  10.00  &  20.18  &  8.40e-04  &  0.59  \\
	\hline
	Argon & $Ar$ &  18.00  &  39.95  &  1.66e-03  &  0.79  \\
	\hline
	Xenon & $Xe$ &  54.00  &  131.30  &  5.46e-03  &  1.36  \\
	\hline
	Air & $N_{2}(76.6)+O_{2}(23.3)$ &  14.45  &  28.90  &  1.20e-03  &  0.71  \\
	\hline
	\end{tabular}
	\caption{Foil material and gap material properties: $Z$, $A$, $\rho$ and $\omega_{P}$ at 101.325 kPa and 293.15 K.}
	\label{tab:material}
	\end{table*}

\end{appendix} 

\end{document}